\documentstyle[aps,preprint,epsf]{revtex}

\begin{document}

\draft
\tightenlines

\title{Absorbing-state phase transitions in fixed-energy sandpiles}
			    
\author{Alessandro Vespignani$^{1}$, Ronald Dickman$^{2}$, 
Miguel A. Mu\~noz$^{3}$,and Stefano Zapperi$^{4}$}

\address{
$^1$ The Abdus Salam International Centre for Theoretical Physics (ICTP)\\
P.O. Box 586, 34100 Trieste, Italy\\
$^2$ Departamento de F\'{\i}sica, ICEx,
Universidade Federal de Minas Gerais,
Caixa Postal 702,
30161-970 Belo Horizonte, MG, Brazil\\
$^3$ Institute {\it Carlos I} for Theoretical and Computational Physics\\
and Departamento de Electromagnetismo y F{\'\i}sica de la Materia\\
18071 Granada, Spain.\\
$^4$PMMH - Ecole de Physique et Chimie Industrielles,
10, rue Vauquelin, 75231 Paris CEDEX 05, France \\
}
\date{\today}

\maketitle
\begin{abstract}
We study sandpile models as closed systems, with conserved  energy density 
$\zeta$ playing the role of an external parameter. 
The critical energy density, $\zeta_c$, marks a nonequilibrium 
phase transition between active and absorbing states. 
Several fixed-energy sandpiles are studied in extensive
simulations of stationary and transient properties, as well as the
dynamics of roughening in an interface-height representation.
Our primary goal is to identify the universality classes of
such models, in hopes of assessing the validity of two recently proposed
approaches to sandpiles: a phenomenological continuum Langevin description
with absorbing states, and a mapping to driven
interface dynamics in random media.  Our results strongly suggest that
there are at least three distinct universality classes for sandpiles.
\end{abstract}



\date{\today}

\section{INTRODUCTION}

Sandpile models \cite{btw} are one of the simplest 
examples of avalanche dynamics, a phenomenon of 
growing experimental and theoretical interest. 
In these models, grains of ``energy'' (sand) are injected 
into the system, while open boundaries \cite{btw} allow 
the system to reach a stationary state, in which energy
inflow (a kind of external drive) and outflow (dissipation) balance.
In the limit of infinitely small external driving,
the system displays a highly fluctuating, 
scale-invariant avalanche-like response: 
the hallmark of criticality.

Ten years after the introduction of the first sandpile automaton by Bak, 
Tang and Wiesenfeld (BTW)\cite{btw}, our understanding of its critical 
behavior remains frustratingly limited, although several variants
of the original model have been studied intensively
\cite{manna,zhang1,directed,tadic}. 
Despite some remarkable exact results \cite{dhar,priezz}, 
and various renormalization group analyses \cite{rg,rg2,rg3}, 
the tempting possibility of assigning these models their 
proper universality classes remains unfulfilled.  
Theoretical and numerical difficulties have
likewise hampered the 
precise estimation of critical exponents.
Only recently was the upper critical dimension $d_c=4$ established
under some assumptions for  the avalanche structure \cite{priezzdc}.

Originally, sandpile models were proposed as the 
paradigm of self-organized criticality (SOC)\cite{btw}, i.e.,
evolution to a critical state 
without tuning of parameters. For this reason, sandpile models were
considered for a long time to inhabit a different world than that of 
standard critical phenomena. Later,
several authors pointed out that, in fact, the SOC state 
can be ascribed to the presence of two infinitely separated time 
scales\cite{hwa,grin,sor95,vz}. The two time scales correspond to the external 
energy input or driving, and the microscopic evolution (``avalanches"). 
This time-scale separation 
(also called {\em slow driving}),
effectively tunes the system to its critical point.
What is the relation between critical states due to infinite 
time-scale separation 
and regular critical points? This question stimulated 
many theoretical studies aimed at elucidating the links among 
sandpile automata 
and models exhibiting nonequilibrium phase transitions, such as 
systems with absorbing states\cite{dvz,vdmz}, 
interfaces in disordered media\cite{midd,pacz,lau,ala}, 
the voter model\cite{directed}, and branching processes\cite{zls}.

In order to make the connections with other
nonequilibrium phenomena  more firm, and to establish universality classes, 
precise critical exponent values are needed. 
Unfortunately, critical exponents governing the deviation from criticality 
cannot be measured in slowly driven sandpiles, which are posed by
definition at their critical point\cite{fast}. 
Thus correspondences between sandpiles and other nonequilibrium 
phase transitions can be only partial and inconclusive.
In order to overcome this conceptual difficulty, a different approach to 
sandpiles has been recently pursued\cite{dvz,vdmz,cmv,mtak}.
It consists in analyzing sandpiles with {\em fixed energy}\cite{tb88}, that is,
in considering the same microscopic rules that define sandpile dynamics, but 
without driving and boundary dissipation. 
In this way the system is closed and thus the
total energy is a conserved quantity, fixed by the 
initial condition, and can be identified as
a (temperaturelike) control parameter.
The system turns out to be critical only for a particular
value of the energy density (equal to that of the stationary,
slowly driven sandpile) and
it is thus possible to study deviations from criticality. 
This approach to sandpiles suggests further analogies with 
systems with absorbing states\cite{reviews} and interfaces
in disordered media \cite{lesch,fisher}. 

The stationary state of standard sandpile models is reached through the 
balance between the input and loss processes, identified by 
the energy addition and dissipation rates $h$ and $\epsilon$, respectively. 
Critical behavior is observed in the slow driving 
regime, in which  the parameters $h$ and $\epsilon$ are 
tuned to their critical values ($h\to 0$ and $\epsilon\to 0$, with
$h/\epsilon \to 0$)\cite{vz,dvz}. In this regime, the system jumps 
among absorbing configurations (in which activity is null) 
via avalanche-like rearrangements. Evidently, in absence of external driving, 
any sandpile model
can fall into an absorbing configuration. The connection to absorbing 
state phase transitions is made more clear by defining {\it closed}, 
fixed-energy sandpiles
in which $h\equiv 0$ and $\epsilon\equiv 0$, and periodic boundary
conditions are imposed. Since the dynamics admits neither input nor loss,
the total energy  $E$ is conserved,
and the energy density $\zeta = E/L^d $
is a tuning parameter. In this case, if the energy density 
$\zeta$ is large enough, the system reaches a 
stationary state with sustained activity, i.e., it is in the 
{\em active} phase\cite{dvz,reviews}. 
On the contrary, for small energy values, the system 
relaxes with probability one into a frozen  configuration, i.e., it 
is in the {\em absorbing} phase.
Separating these two regimes is a critical 
point ($\zeta = \zeta_c$) with marginal
propagation of activity. 

Once it is appreciated that fixed-energy sandpiles exhibit a continuous
transition to an absorbing state,
the existence of a critical stationary state in
the corresponding driven dissipative sandpile is easily understood.
That is because
energy is added only in the absence of activity ($\zeta < \zeta_c$) while
dissipation occurs only in the presence of activity ($\zeta > \zeta_c$).
Thus $d\zeta/dt$ is positive for $\zeta < \zeta_c$, and vice-versa, leaving
$\zeta_c$ as the only possible stationary value for the energy 
density\cite{inf}.
(The condition that dissipation and hence activity be absent in the subcritical
phase makes the absorbing nature of this phase an essential ingredient of SOC.)
Since SOC means tuning a system to its critical point by means of an 
infinitely slow drive, it is natural to try to understand 
the critical behavior 
first in the simpler context of a fixed-energy model.  But while many examples 
of absorbing-state phase transitions 
have been studied in detail in recent years, we will see that characterizing
sandpile criticality, even in the fixed-energy formulation, is a 
nontrivial project.

In this paper we define and study {\em fixed-energy sandpiles} (FES)
with various microscopic dynamics. In particular we analyze the 
BTW sandpile \cite{btw}, the stochastic Manna model\cite{manna,mannadhar},
and a model with random mixing
of a (real-valued) energy: the shuffling model \cite{zhang2}
(full definitions are given in the following section). 
We show that all of these models exhibit an absorbing-state phase transition
at a critical value $\zeta_c$ of the energy density.  
What distinguishes the sandpile from other models with absorbing states
is that the control parameter $\zeta$ represents the global value 
of a conserved field. This phase transition
is the basis of the critical behavior of 
driven self-organized sandpiles. 
The transition is also studied using mean-field approximations,
which yield good qualitative predictions for the order parameter and 
transition points.

Using the insights provided by the connection with absorbing states,
we discuss in detail the attempt to construct a field theory 
for sandpiles\cite{vdmz}. The latter is a generalization of Reggeon
field theory (RFT)\cite{rft}, the minimal continuum theory describing
absorbing-state phase transitions\cite{bs94}. We also discuss an alternative 
approach that considers sandpiles 
from the perspective of linear interface models (LIM) in disordered 
media\cite{midd,pacz,lau}. Since continuum descriptions have proved to be 
of fundamental importance in understanding universality and critical behavior,
we analyze in detail open questions and possible  
improvements of these theoretical approaches.  

For all the models mentioned, we report results of
simulations close to the critical point, and discuss them in terms 
of universality classes.
Numerical results indicate three distinct critical behaviors, depending
upon the microscopic dynamics of models. In particular, the BTW model 
defines a  critical behavior {\it per se}, related to the 
deterministic nature of the dynamics. We 
find striking evidence of non-ergodicity in 
the BTW FES: an anomalous transient to the stationary 
state, and lack of self-averaging.
Stochastic automata, such as the  Manna model, have a critical behavior 
that is rather close to the one of linear interface depinning models.
Finally, the shuffling model shows a critical behavior that could be 
compatible with the RFT universality class. However, the nonlocal dynamics 
of this model merits a detailed examination. 
It is also important to note that all models show a violation of
certain scaling relations usually associated with absorbing-state phase transitions.
This seems to point out the particular role of the conserved field in these
systems. Finally, we discuss the numerical results in the perspective of 
the theoretical frameworks mentioned above.

The outline of this paper is as follows:
after defining the models in Sec. II, we
discuss the generalized RFT theory (Sec. III) and LIM approach (Sec. IV) to 
FES models. We analyze from a critical perspective the approximations and 
hypotheses involved in these approaches. In particular, 
we discuss the nature of the different noise terms; this turns out to be 
essential to the identification of universality classes. 
In Sec. V we present the results of extensive simulations 
in two dimensions, and analyze them
in the perspective of absorbing-state transitions
\cite{dvz,vdmz}, and the LIM mapping, which focuses on 
the roughness of a suitably defined interface \cite{midd,pacz,lau}.
We find differences between BTW, Manna and fully stochastic FES exponents that
persist upon enlarging  the system size. 
Sec. VI is concerned with the origins of these differences and possible
improvements in the theoretical descriptions to capture the true critical 
behavior of FES models.  A brief summary is provided in Sec. VII.
Mean-field theory approaches at the one- and two-site levels are described in the
Appendix.

\section{Fixed-energy sandpiles}

In this paper we consider three different sandpile models.
All are defined on a $d$-dimensional
hypercubic lattice ($d=2$ in this study); the
configuration is specified by giving the {\it energy},
$z_i$, at each site.  The energy may take integer or real values, depending on the model,
but is nonnegative in all cases.  The specific models are defined as follows.
\vspace{.5cm}

\noindent {\it BTW model} \cite{btw}: Each active site, i.e.,  with (integer) energy
greater than or equal to the {\em activity threshold} $z_{th}$ ($z_i \geq z_{th} = 2d$), 
topples at unit rate, i.e., $z_i \rightarrow z_i - z_{th}$, and $z_j \rightarrow z_j + 1$ at
each of the $2d$ nearest neighbors of $i$.   The toppling rate is introduced in
order to define a Markov process with finite transition rates between configurations
that differ at a small number of sites.
The next site to topple is selected at random from the set of active sites; this is
the only stochastic element in the dynamics.  (The initial configuration is, 
in general, random as well.)  The BTW dynamics with {\it parallel} updating
(all active sites topple at each update) is completely deterministic, 
and it has been possible
to obtain many exact results for the driven sandpile in this case, due to
the Abelian property\cite{dhar}. This property implies that the order
in which active sites are updated is irrelevant
in the generation of the final (inactive) configuration. 
Accordingly, it is reasonable to expect that 
sequential or parallel updating does not affect the qualitative 
behavior.  The BTW model is the prototypical sandpile model, and has been the subject of 
extensive numerical studies \cite{grasma,manna2,lubeck1}. 
Despite the huge numerical effort devoted to the analysis of its 
critical behavior, the model presents scaling anomalies which have precluded 
a definitive characterization. The scattered numerical values of 
the avalanche critical exponents were recently interpreted in terms
of multiscaling properties \cite{stella}.
\vspace{.5cm}

\noindent {\it Manna sandpile} \cite{manna,mannadhar}: 
In this case  $z_{th} =2$ regardless of the
number of dimensions; the energy is again integer-valued.
The two particles liberated when the site $i$ topples 
move {\it independently} to randomly chosen 
nearest neighbors $j$ and $j'$ (That is, $j=j'$ with 
probability $1/2d$) \cite{inex}.
 This model has a stochastic dynamics, which still enjoys a ``stochastic''
Abelian property, as shown recently by Dhar \cite{mannadhar}. The
Manna model has also been the subject of many numerical studies. 
Together with the BTW model, it has been at the center of the long 
debate over universality classes for (driven) 
sandpiles \cite{ben,csvz,lubeck2}, 
that we will discuss in 
later sections. The Manna model, fortunately, has a regular 
scaling behavior. The most recent analyses provide a coherent picture of 
its critical properties and exponent values\cite{csvz,lubeck2,cvz,romu}.
\vspace{.5cm}

\noindent  {\it Shuffling model}\cite{zhang2} : This  model has nonnegative 
real-valued energies. 
When a site $i$ topples, the energy $Z = z_i + \sum_{j NN i} z_j$ at that site
and its nearest neighbors is redistributed randomly amongst these five sites.  
That is, we generate random numbers $\eta_1,...\eta_5$, uniform on [0,1],
and let $z_j \rightarrow z'_j =
\eta_j Z/(\eta_1 + \cdots + \eta_5)$ ($j = 1,...,5$).
Sites with energy $z'_j \geq z_{th} = 2$ topple with probability one. 
In addition, the nearest neighbors of the toppling
site that have energy $z'_j < z_{th}$ also become active
with probability $z'_j/z_{th}$. This model contains 
stochasticity  in each ingredient of the dynamics, and for this reason can be considered a 
fully stochastic model.  
It is clearly non-Abelian: the final configuration 
depends dramatically upon the order in which sites are updated.
The parallel-updating version studied in this work exhibits an 
interesting nonlocal dynamical effect. At each update, the energy
around a site is shuffled among nearest-neighbor sites. If a nearest-neighbor (or
next-nearest neighbor) pair of sites
are both active, the energy at a certain site or sites will be shuffled 
twice within a single time step.  For larger aggregates of active sites, the 
reshuffling may involve the same site several times.
In particular, energy can be transported 
over large distances by consecutive shuffling events along the front 
of active sites. This non-locality will create a mixing effect in the 
energy transport that one expects to influence the critical behavior.

In the present paper, we study the Manna and shuffling models with the parallel 
updating customarily used in sandpile automata. The BTW model is implemented 
using random sequential dynamics, with each active site having
a toppling rate of unity.  The next site to topple is chosen
at random from a list of active sites, which must naturally be updated following
each toppling event.  The time increment associated with each such event
is $\Delta t = 1/N_A$, where $N_A$ is the number of active sites.
This is the mean waiting-time to the next event, if we were to choose sites blindly, instead of using a list.  (In this way, $N_A$ sites topple per unit time, just as in a
simultaneously updated version of the model.) Since the BTW model is 
Abelian, the choice of updating (parallel versus sequential) should be irrelevant to the asymptotic 
critical properties. This has been tested in independent 
simulations using parallel dynamics\cite{test}.

In a FES, the energy density $\zeta$ is fixed in the initial condition. The latter is generated by
distributing $\zeta L^d$ particles randomly among the $L^d$ sites,
yielding an initial (product) distribution that is spatially homogeneous and uncorrelated.
Once the particles have been placed the dynamics begins. The condition to have at least one active site in the initial configuration is trivially satisfied on large lattices,
for the $\zeta$ values of interest, i.e., close to the critical value.
(For large $L$, the initial height at a given site is essentially a 
Poisson random variable, and the probability of having no active sites
is exponentially decreasing with the lattice size). 
It is worth remarking that while the initial conditions are statistically
homogeneous, the energy density is not perfectly smooth. For $ 1 \ll l \ll L$, the
energy density on a set of $l^d$ sites is essentially a
Gaussian random variable with mean $\zeta$ and variance $\sim l^{-d}$.
The initial value of the critical-site density $\rho_c$ (sites that 
become active upon receiving energy), moreover, is generally far from its stationary value, complicating relaxation to the steady state.

If after some time the system falls into a configuration with 
no active sites, the dynamics is permanently frozen, i.e., the 
system has reached an absorbing configuration. We shall see that as we vary $\zeta$, 
fixed-energy sandpiles show a phase transition separating an absorbing phase (in which  
the system always encounters an absorbing configuration), from an active
phase possessing sustained activity \cite{immortals}. This is a continuous 
phase transition, at which the system shows critical behavior.  The order
parameter is the stationary average density of active sites $\rho_a$, which
equals zero for $\zeta<\zeta_c$, and follows a power 
law, $\rho_a\sim(\zeta-\zeta_c)^\beta$, for $\zeta>\zeta_c$.
The correlation length $\xi$ and relaxation time 
$\tau$ both diverge as $\zeta\to\zeta_c$; their critical behavior is 
characterized by the  exponents $\nu_\perp$ and $\nu_\parallel$, defined 
via $\xi\sim|\zeta-\zeta_c|^{-\nu_\perp}$ and 
$\tau\sim|\zeta-\zeta_c|^{-\nu_\parallel}$, respectively.
The dynamical critical exponent is defined via $\tau\sim\xi^z$,
which implies $z=\nu_\parallel/\nu_\perp$. The exponents $\beta$, $\nu_\perp$
and $\nu_\parallel$
define the stationary critical behavior at the absorbing-state phase transition
\cite{reviews}. In the vicinity of the critical point,
where $\xi$ is very large, the actual characteristic length of the system
is the lattice size $L$. We shall see that the application of finite-size scaling allows 
us to locate the critical point as well as estimate 
critical exponents. 

\section{Sandpiles as systems with absorbing states}

In this section we discuss a recently proposed
phenomenological field theory of sandpiles \cite{vdmz}.
Our main goal is to clarify the connection
between fixed-energy sandpiles and Reggeon field
theory (RFT), which is the minimal field theory describing 
absorbing-state phase transitions\cite{rft,bs94} 
(whose prototypical examples are directed percolation (DP)\cite{dp} 
and contact processes (CP)\cite{cp}).

In Ref.~\cite{vdmz} we proposed a Langevin description
for sandpiles by considering  the mean-field description of 
sandpiles reported in Ref.\cite{vz,dvz}, and introducing spatial
dependence and fluctuations. This allows a derivation that is 
based on the microscopic dynamics of  sandpile automata, but 
involves several approximations.

Here we show how to write down a general Langevin description of sandpiles by using very general symmetry considerations. This results in a complete description, but one
that is not easy to deal with, unless the proper approximations are 
introduced. After the introduction of some specific assumptions
regarding noise terms, we recover the results of Ref.~\cite{vdmz}. 
On the other hand, the present more general treatment indicates possible modifications 
that may be needed for a complete
characterization of sandpile models.

In sandpiles, the order parameter is $\rho_a$, 
the density of active sites (i.e., whose height 
$z \geq z_c$) \cite{vz,dvz,tb88}; if at a given time $\rho_a({\bf x})=0$ 
for all {\bf x}, the system 
has reached an absorbing configuration. The only dynamics in the model
is due the field $\rho_a({\bf x})$, 
which is coupled to the local energy density, 
$\zeta({\bf x},t)$, 
which enhances or depresses the generation of new active sites\cite{notarft1}. 
We therefore consider the dynamics of the local order-parameter field
$\rho_a({\bf x},t)$ in a coarse-grained description, bearing in mind that
the energy density $\zeta({\bf x},t)$ is a {\it conserved} field.
Note that both $\rho_a({\bf x},t)$ and $\zeta({\bf x},t)$ are nonnegative.
The most general dynamical equation that imposes local conservation of energy is
\begin{equation}
\frac{\partial \zeta({\bf x},t)}{\partial t} =\nabla^2( 
f_\zeta[\{\rho_a\},\{\zeta\}])
+\nabla \cdot [g_\zeta(\{\rho_a\},\{\zeta\})\overrightarrow{\eta}({\bf x},t)],
\label{conservation}
\end{equation}
where $f_\zeta$ and $g_\zeta$
are functionals of $\rho_a$ and $\zeta$. Conservation is enforced 
by the $\nabla^2$ term and the standard form of 
conserving noise, as
for example in Cahn-Hilliard-type equations \cite{Bray} 
($\overrightarrow{\eta}$ is a $d$-component vectorial noise).
The dynamical equation for the density of 
active sites can be written analogously as 
\begin{equation}
\frac{\partial \rho_a({\bf x},t)}{\partial t} = 
f_a(\{\rho_a\},\{\zeta\})
+g_a(\{\rho_a\},\{\zeta\})\eta({\bf x},t),
\label{act1}
\end{equation}
where $f_a$ and $g_a$
are functionals of $\rho_a$ and $\zeta$ and $\eta({\bf x},t)$
is an uncorrelated Gaussian noise. We note that
$\eta$ is a {\it nonconserved} noise: the active-site density
is not a conserved quantity.  
The functionals $f_a$ and $f_\zeta$, and variances $g^2_a$ and $g^2_\zeta$ 
appearing on the r.h.s. of Eqs. (\ref{conservation}) and (\ref{act1}) are
analytic functions (polynomials) of the local densities 
and (in principle) their spatial derivatives.

The right-hand-sides of Eq.s~(\ref{conservation}) and (\ref{act1}) 
must vanish when $\rho_a = 0$  
(if they did not, the state $\rho_a = 0$ would not be
absorbing!).
This implies that none of the functionals $f_a$, $g^2_a$, $f_\zeta$, 
and $g^2_\zeta$ contain terms independent of $\rho_a$;
they are functions of $\rho_a({\bf x},t)$ and the product 
$\zeta({\bf x},t)\rho_a({\bf x},t)$\cite{reviews}. 
In this way activity is sustained only if $\rho_a({\bf x},t) > 0$. 
It is convenient at this point to introduce a reference
value $\zeta_0$ of $\zeta$ (for instance the global average energy), 
and expand the term $\propto \zeta \rho_a $ about $\zeta_0$. Introducing
$\Delta \zeta ({\bf x},t) \equiv \zeta ({\bf x},t) - \zeta_0$
we can express all the functionals as functions of
$\Delta\zeta({\bf x},t)\rho_a({\bf x},t)$, where all 
terms of the form $\zeta_0 [\rho_a({\bf x},t)]^n$
are absorbed into the coefficient of $[\rho_a({\bf x},t)]^n$, 
$\zeta_0$ being constant.

In order to write the various functionals more explicitly, we have 
to consider the symmetry of the lattice in question. For 
isotropic  models the system is inversion-symmetric under 
${\bf x}\to-{\bf x}$, so that odd powers of gradients, such as 
$\nabla\rho_a$, are forbidden. This leaves us with functionals such as
\begin{equation}
f_a(\{\rho_a\},\{\zeta\})= D_a\nabla^2\rho_a({\bf x},t)-
r\rho_a({\bf x},t)+\mu\rho_a({\bf x},t)\Delta \zeta ({\bf x},t)
-b\rho_a^2({\bf x},t)+.~.~.~.
\end{equation}
where $D_a, r, \mu$ and $ b$ are constants whose connection with 
the microscopic dynamics will be clarified below. The functionals 
$f_\zeta$, $g_a$ and $g_\zeta$ have similar forms.
If we do not want to deal with an infinite set of 
power and derivative terms in $\rho_a({\bf x},t)$ and 
$\Delta \zeta ({\bf x},t)$, we have to identify the relevant 
terms from the renormalization group point of view. This can be done via
power counting analysis at the upper critical dimension.
This implies the knowledge of the noise term, i.e., we have to decide
the terms to retain in $g_a$ and $g_\zeta$.
The most relevant term is the linear one, corresponding to
$g_a\sim g_\zeta\sim\rho_a^{1/2}({\bf x},t)$ \cite{rft,reviews}.
In RFT, the rationale for the noise variance being proportional
to the local order parameter is that the numbers of elementary
(birth and death) events in a given space-time cell 
are Poissonian random variables, so the variance is equal to 
the expected value.
That the noise term for sandpile models has the same form as in RFT
is by no means guaranteed. For instance, the BTW model is 
fully deterministic, and 
the nontrivial assumption that at the coarse-grained level it 
is described by a time-dependent noise should be tested. 
Further, the fact that the 
field $\zeta({\bf x},t)$ is conserved could affect the noise form. 
In fact, it is well known 
that additional symmetries on the fields can change the 
noise form\cite{cardy}. In the absence of an exact derivation of the 
noise terms, we proceed by showing the Langevin description
resulting from the choice of a RFT-like noise. 

Assuming  RFT-like noise terms, the activity equation takes the form
\begin{eqnarray}
\frac{\partial \rho_a({\bf x},t) }{\partial t} & =&
D_a\nabla^2 \rho_a({\bf x},t) - r \rho_a({\bf x},t)   
 -b \rho_a ^2({\bf x},t)                               \nonumber \\
&+& \mu \rho_a({\bf x},t) \Delta \zeta ({\bf x},t)
+ \eta_a({\bf x},t)  ,
\label{active}
\end{eqnarray}
where $\eta_a=\rho_a^{1/2}\eta$. Here we have retained only 
relevant terms with respect to the noise considered.
In mean-field theory the critical
point corresponds to $r = r_c= 0$; we expect fluctuations to renormalize 
$r_c$ to a nonzero value.
In any case, the value of $r$ depends on $\zeta_0 $, i.e.,
the energy density $\zeta_0$ plays the role of a
(temperaturelike) control parameter.

The evolution of $\Delta \zeta ({\bf x},t) $ is governed 
only by the most relevant term in the functional $f_\zeta$, 
that is, the one linear in $\rho_a$. The equation 
may be integrated formally to yield
\begin{equation}
\Delta \zeta ({\bf x},t) =  \Delta \zeta ({\bf x},0)  +
\int_0^t  d t' \left[ D_{\zeta} \nabla^2 \rho_a ({\bf x},t') +
\nabla \cdot \left(
\sqrt{\rho_a({\bf x},t')}{\overrightarrow{\eta}} \right) \right].
\label{rhoc} 
\end{equation}
Substituting this into Eq. (\ref{active})
and disregarding irrelevant higher order terms, 
the proposed Langevin equation for fixed-energy sandpiles 
becomes \cite{vdmz}:
\begin{eqnarray}
\frac{\partial \rho_a ( {\bf x},t)}{\partial t}
& =& D_a\nabla^2 \rho_a({\bf x},t) - r ({\bf x}) \rho_a({\bf x},t)
 - b \rho_a^2({\bf x},t)   \nonumber \\
& + & w \rho_a({\bf x},t) \int_0^t dt' \nabla^2 \rho_a({\bf x},t')
  + \sqrt{\rho_a} \eta ( {\bf x},t).
\label{main}
\end{eqnarray}
$\eta$ is a Gaussian white noise whose only non-vanishing cumulants are
$\langle \eta(  {\bf x},t) \eta ( {\bf x'},t')\rangle = D \delta({\bf x-x'})
\delta(t-t')$, $c, b$ and $w$ are fixed parameters, and
the coefficient of the linear term, 
\begin{equation}
r({\bf x}) =r -\mu \Delta \zeta ({\bf x},0),
\label{quench}
\end{equation}
inherits its spatial dependence from the initial
energy distribution $\Delta \zeta ({\bf x},0)$.
Observe that $b$ has to be positive to ensure stability; $w>0$ follows from
the diffusion coefficient $D_{\zeta} > 0 $.
This equation recovers the result obtained in Ref.~\cite{vdmz};
we refer the reader interested in a more phenomenological
approach to that paper.

We find, by standard power-counting analysis, that the upper
critical dimension of this theory is $d_c=4$ \cite{future}. Above $d_c$,
a qualitatively correct mean-field description
is obtained by dropping the noise and gradient terms
and replacing $\zeta({\bf x},0)$ by the spatially uniform $\zeta=\zeta_0$, yielding:
\begin{equation}
\partial_t \rho_a(t)=-\overline{r}\rho_a(t) 
-\overline{b}\rho_a^2(t).
\end{equation}
The critical point, $\zeta=\zeta_c$,
corresponds to  $\overline{r}=0$.  
Above $\zeta_c$, we have an active stationary state
with $\rho_a \sim (\zeta-\zeta_c)^{\beta}$ with $\beta=1$; 
for $\zeta < \zeta_c$,
the system falls into an absorbing configuration in which $\rho_a=0$.
Other MF critical exponents can be calculated as well. 

The present Langevin equation resembles RFT,
except for the spatial dependence of $r$ and the non-Markovian term.
Both stem from the interaction between activity 
with the energy background.
Let us present here some comments on these two extra terms.


{\em Quenched disorder}: in the absence of the memory term, 
and for generic initial conditions, 
$\Delta \zeta ({\bf x},0) \neq const.$,
Eq. (\ref{main}) is the field theory
of directed percolation with {\it quenched disorder}. 
 Disorder
is known to be a relevant perturbation in DP below 
$d_c = 4$ \cite{noest,dcp1,noest88,dcp2,madcp}.
On the other hand, the memory and spatially-dependent linear terms
{\it together} represent coupling to the energy density, which is not
quenched-in, but relaxes via the diffusion of activity [see Eq. (\ref{quench})].
Thus the effect of a spatially-dependent $r$, in the present context,
is not that of quenched disorder.  
In fact, we expect the physical effects of quenched disorder,
and the present coupling to a conserved energy density (frozen
only in the absence of activity), to be quite different\cite{notarft2}.
A handwaving argument to justify this assertion is the following:
In the active stationary state, close to the critical point,
activity will tend to be localized at any given moment,
and a given point {\bf x} will experience 
bursts of activity interspersed amongst dormant intervals.
As activity alternately enters and vanishes from the neighborhood of
{\bf x}, the positive and negative contributions to the Laplacian
memory term in Eq. (\ref{main}) will largely cancel, and so this
term will be dominated by the most recent changes in the state of the region.
Thus the initial spatial variation in $r({\bf x},0)$ will effectively
be forgotten in the stationary state.
Another way to see this is to note that the effects of quenched disorder are
found in a {\em non-Markovian} version of the contact process \cite{madcp,janssen} (using the so-called
``run-time statistics") in which the creation rate at site $i$ is
$\lambda_i (t) = (c_i + a)/(n_i + a + 1)$, where $a$ is a parameter, and $c_i$
represents the number of creation events out of $n_i$ total events at site $i$,
up to time $t$.  Evidently, sites which by chance have enjoyed a larger fraction
of creation events in the past are likely to continue to do so, mimicking a
quenched random creation rate.  In the present case, the effective
creation rate ($\propto -r(x)$) is 
$\lambda(x,t) = \lambda(x,0) + w \int_0^t dt' \nabla^2 \rho_a (x,t')$.
Now, regions with $\rho_a$ larger than $<\rho_a>$ tend to 
have $\nabla^2 \rho_a < 0$.
Thus the non-Markovian term provides
a stabilizing, negative feedback on the creation rate.  Regions
currently experiencing above-average activity will be harder to excite in the
near future.  (Note however, that $\int r(x,t) dx $ is time independent, since
$\int \nabla^2 \rho_a dx = 0$.)
While the non-Markovian term effectively erases the initial distribution
$r(x,0)$, we do expect the spatial dependence of $r$
to play an important role when we consider avalanches, i.e., the
spread of activity from a localized seed, in a nonuniform energy density.

{\em Non-Markovian term}: 
 As we have just discussed, this term enables the theory to 
forget the quenched, stochastic reproduction rate $r({\bf x},0)$.
Naively, its associated coefficient, $w$,
has the same dimensionality as $b$ and $D$, which are the
two marginal parameters of RFT at its upper critical dimension, $d_c=4$.
Below $d_c$ 
we expect the critical fixed point to be renormalized to
$r=r^*$, defining a renormalized $\zeta_c$ and nontrivial critical exponents.
If the non-Markovian term is irrelevant, the field theory would be governed 
at criticality by the RFT fixed point. 
In $d=2$ the RFT critical behavior is characterized by 
$\beta\simeq0.58$, $\nu_\perp\simeq0.73$ and 
$z\simeq1.77$\cite{reviews}.
We shall see in the following sections that numerical results
are not compatible with this picture in the BTW and Manna case.
This calls for a full RG 
analysis  of Eq.~(\ref{main}). Unfortunately, this is 
a very dificult task because of primitive 
divergencies appearing in the perturbative approaches.
A discussion of the RG treatment of the present field theory 
will be reported elsewhere\cite{future}.

Possible modifications and generalizations of Eq.~(\ref{main}), and 
their implications for critical behavior, 
will be discussed in later sections. Finally,  
a microscopic derivation of the field theory would ensure that the
conservation symmetry has been properly taken into account in the 
present phenomenological approach.

\section{Sandpiles as interfaces in random media}

A connection between sandpiles and interfaces moving 
in disordered media can be obtained by defining a variable
$H(i,t)$ that counts the number of topplings (instances of activity) 
at site $i$ up to time $t$. This variable defines a growing surface
in a $d+1$ dimensional space. The interface is said to be 
in the pinned phase if its disorder-average
velocity $<\!\partial_t H(i,t)\!>$ is null;
a finite velocity marks the moving
phase. It is then easy to recognize that the pinned phase in 
interface models is completely analogous
to an absorbing state, while the moving phase corresponds to 
an active state \cite{alonnote}. 
To make this correspondence more precise let us note that 
a nonzero interface velocity is only possible if active sites
are present in the system; equivalently we can notice
that $\partial_t H(i,t)=\rho_a(i,t)$, 
so in either representation the dynamically
active phase is restricted to the
regime with nonvanishing $\rho_a(x,t)$.
In this way it is evident that 
pinned (unpinned) and absorbing (active) states are just two ways of 
looking at the same physical situation.
The connection between driven sandpiles and interfaces 
was first proposed by Narayan and Middleton and by Paczuski and Boettcher
\cite{midd,pacz} and recently generalized by
Lauritsen and Alava \cite{lau,ala}  who  provided
a direct mapping between the BTW model and a linear interface 
with quenched disorder.  In the following we adapt their
approach to fixed-energy sandpiles.

Let $H_i(t)$ be the number of topplings 
at site $i$ up to time $t$, and $z_i(t)$ the energy at $i$ at time $t$.
The latter is evidently the difference between the inflow
and the outflow of energy at site $i$ in the past.
The outflow is given by $2dH_i(t)$, since in each toppling $2d$ 
particles are expelled from the site. 
There are two contributions to the inflow,
the first being the energy $z_i(0)$ present at time $t=0$.
The second comes from topplings of the nearest-neighbor sites, and 
can be expressed as $\sum_{NN} H_j(t)$. 
Summing the above contributions we obtain:
\begin{eqnarray}
z_i(t)&=&z_i(0)+\sum_{j NN i} H_j(t)-2dH_i(t)\\
&=&z_i(0)+\nabla_D^2 H_i(t),
\label{hdiscr}
 \end{eqnarray}
where $\nabla_D^2$ stands for the discretized Laplacian.

Since sites with $z_i(t) > z_c = 2d-1$ topple at unit rate, 
the dynamics of the height follows

\begin{equation}
\frac{d H_i(t)}{d t} = \Theta [z_i(0) + \nabla_D^2 H_i(t) - z_c],
\label{hdyn}
\end{equation}
where $d H_i(t)/d t$ is a shorthand notation for the {\it rate} at
which the integer-valued variable $H_i(t)$ jumps to $H_i(t) +1$, and
$\Theta (x) = 1$ for $x > 0$ and is zero otherwise.
Since $z_i(t)$ takes integer values, the smallest
argument of the $\Theta$-function yielding a nonzero toppling rate is unity.
If we replace $\Theta (x)$ by $x$, and assume this change to be
irrelevant for critical properties \cite{nota_dis}, then the
BTW FES is mapped onto a discretized  
Edward Wilkinson (EW) equation \cite{lesch,barabasi} with quenched
disorder, represented by the fluctuations in the $z_i(0)$ term.
A noise term of this kind, which
varies from site to site, but is time-independent, is referred to as 
{\it columnar noise} in the field of interface dynamics \cite{parisi,nota_lau}.

To understand the phenomenology of Eq.~(\ref{hdyn}), let us define 
the average initial energy as $f=\langle z_i(0) \rangle$. 
There are three different possibilities. 
\begin{itemize}
\item [1)] If $f$ is small then with probability one the system is eventually pinned by 
disorder.
\item[2)] If $f$ is large enough, the system has a finite velocity and 
keeps moving indefinitely. 
\item[3)] Separating these two regimes is a critical point 
marking the depinning transition.
\end{itemize}
Thus the phase transition in the BTW FES is analogous to a depinning 
transition. If the caveat noted above regarding the replacement 
$\Theta(x)\to x$ turns out to be unimportant, then the transition 
should show the same scaling properties as depinning in the 
Edward-Wilkinson equation with columnar noise. 

How are these results changed for the Manna model?  For the outflow at
site $i$ we now have $2H_i(t)$, since only two particles are transferred 
in each toppling event.  The total input is the sum of the initial energy, 
$z_i(0)$, and a
{\it stochastic} contribution $I_i(t)$ associated with topplings at the nearest
neighbors of $i$:

\begin{equation}
I_i(t) = \sum_{j NN i} \sum_{\tau = 1}^{H_j(t)} \eta_{i,j}(\tau) \;,
\label{manin}
\end{equation}
where the $\eta_{i,j}(\tau)$
are a set of independent (for $i$ fixed!), identically distributed 
random variables that specify
the number of particles  (0, 1, or 2) received by site $i$ at the $\tau$-th 
toppling of site $j$.  Thus
\begin{equation}
\eta_{i,j}(\tau)  =\left\{ \begin{array}{ll}
0 & \mbox{with probability $(1-1/2d)^2$}  \\
1 & \mbox{with probability $(1-1/2d)/d$}  \\
2 & \mbox{with probability $(1/2d)^2$}  
\end{array} \right.
\label{etadist}
\end{equation}
Of course, the variables associated with different acceptor sites 
$i$ are highly correlated,
since $\sum_i \eta_{i,j}(\tau) = 2$.
$\eta_{i,j}(\tau)$ has mean $1/d$ and variance $(1-1/2d)/d$.  
It is convenient to introduce $\xi_{i,j}(\tau) \equiv \eta_{i,j}(\tau) - 1/d$, 
which has zero mean, the same variance
as $\eta_{i,j}(\tau)$, and obeys $\sum_i \xi_{i,j}(\tau) = 0$.
We may now write the analog of Eq. (\ref{hdiscr}) for the Manna model:

\begin{equation}
z_i(t) = z_i(0) + \frac{1}{d} \nabla_D ^2 H_{i}(t) +
 \sum_{j NN i} \sum_{\tau = 1}^{H_j(t)} \xi_{i,j}(\tau) .
\label{mandisc}
\end{equation}
To obtain a simple EW-like equation for the height in the Manna model,
we must (1) ignore the correlations between noise terms associated 
with different sites, and (2) imagine that the noise is updated when 
site $i$ itself, rather than one of its
neighbors, topples; we will denote the noise term as $\xi_i(H)$.  
Under these assumptions we may write
\begin{equation}
\frac{d H_i(t)}{d t}  =\left\{ \begin{array}{ll}
1 \;, 
& \mbox{if}\;\; z_i(0) +\frac{1}{d} \nabla_D^2 H_i(t)  + \xi_i(H) \geq 2  \\
0 \;, & \mbox{ otherwise}.  \end{array} \right.
\label{mannalim}
\end{equation}

We have obtained an EW-like equation with {\it quenched} as well as
columnar disorder,
the so-called linear interface model. This last equation has been
studied extensively both theoretically and 
numerically\cite{lesch,fisher,barabasi}. If the previously discussed 
approximations are irrelevant, the Manna model should belong to the 
LIM universality class \cite{lesch,fisher}. The fact that the correlations
between the noise terms are short range argues  in favor of this 
conclusion \cite{ala}.

The shuffling model deserves a  particular note. In fact, 
it is not obvious that we can write an 
exact relation between
the (continuous-valued) energy $z_i(t)$ and the height 
(number of topplings) $H_i(t)$. It is possible to write an interface equation
for the shuffling model if we introduce some 
phenomenological constants and approximations beyond those used in
the Manna case. We do not report the full derivation, 
which finally leads to an equation of 
the form of Eq. (\ref{mannalim}).

We have seen that two issues remain unresolved:

i) Whether the approximations involved in the Manna and shuffling cases 
change the critical behavior from the LIM universality class.

ii) Whether the various  models are in the same universality class, 
since even if the approximations in i) are irrelevant, 
the Manna equation involves quenched
as well as columnar noise, while only the latter appears in the BTW equation.

In order to answer the above questions analytically,  a 
more rigorous study of the noise terms appearing in the interface 
equations is needed. This is analogous to the Langevin description
of the previous section. We caution however that this analogy
does not imply that it is easy, or even possible, to translate equations or
results from one language to the other. 
For example, to the best of our knowledge, no one has succeeded
in writing down an interface-like equation equivalent to RFT\cite{notalim1}.

From a numerical point of view it is possible to measure various exponents
characterizing the behavior of moving interfaces. 
Many of these exponents can be
related to those measured in the context of absorbing-state phase transitions. 
It appears clear from the previous discussion that the driving force
in the interface picture is equivalent to the energy density $\zeta$.
This is the control parameter, and the exponents
$z$ and $\nu_\perp$ are the same in both pictures. Moreover, 
the order parameter exponent $\beta$ is equivalent to the interface 
velocity exponent usually measured in interface depinning models. 
More interestingly, associated with the interface picture are 
new exponents, related to the interface roughness, defined as :
\begin{equation}
W^2(L,t)=\frac{1}{L^d}<\sum_i(H_i(t)-\overline{H(t)})^2>
\label{w2}
\end{equation}
where $\overline{H(t)}=L^{-d}\sum_iH_i(t)$ and the $<>$ brackets represent
an average over different realizations. 
In general one expects $W^2$ to exhibit an $L$-independent, power-law growth 
regime prior to saturating, that is \cite{barabasi}
\begin{equation}
W^2 (t,L)  \sim \left\{ \begin{array}{ll}
t^{2 \beta_W} \;, & t \ll t_{\times}  \\
L^{2 \alpha}\;, & t \gg t_{\times},  \end{array} \right.
\label{w2scal}
\end{equation}
where the crossover time $t_{\times} \sim L^z$.  
The limiting behaviors described
above follow from the dynamic scaling property,

\begin{equation}
W^2(t,L) = L^{2 \alpha} {\cal W}(t/L^z) \;,
\label{wdynsc}
\end{equation}
where the scaling function  
${\cal W}(x) \sim x^{2 \beta_W} $ for small $x$, and attains
a constant value for $x \rightarrow \infty$.  
The dynamic exponent thus satisfies the scaling relation  
$z = \alpha/\beta_W$ (first proposed by Family and Viseck \cite{FV}).
We expect a data collapse for different system 
sizes in a plot of
$L^{-2 \alpha} W^2(t,L)$ versus $t/L^z$.  
The roughness exponents are related via scaling relations to the other 
critical exponents. One may show, for example, that $\beta_W =1-\theta$.
To see this, note that in the power-law growth regime, for which the
correlation length $\xi (t) \ll L$, growth events in different regions are
uncorrelated.  Given the scaling property of the single-site height probability,
$ P[H_i(t)] = f [H_i(t)/\overline{H(t)}]$, we have 
$W^2 (t) = {\rm var}[H_i(t)] \propto [\overline{H(t)}]^2$.
Since $\overline{H(t)}$ is simply
the integrated activity, 
$\overline{H(t)} = \int_0^t dt' \rho_a (t')  \propto t^{1-\theta} $,
yielding $\beta_W = 1 - \theta$.

At this point it is well to raise a caution regarding  
the naive application of 
scaling laws such as those mentioned in the preceding paragraph. 
Recent numerical studies have 
revealed that many growth models may exhibit {\it  anomalous roughening},
i.e., the local width (calculated on `windows' of size $l << L$) scales with an
exponent, $\alpha_{loc}$, other than $\alpha$. In these cases, 
simple scaling a la Family-Viscek does not hold.
Technically this corresponds 
to the following situation:
$ W(l,t) \approx t^{\beta_W} {\cal F}_A(l/t^{1/z})$,
with an anomalous scaling function given by:
\begin{equation}
\label{F_A}
{\cal F}_A (u) \sim
\left\{ \begin{array}{lcl}
     u^{\alpha_{loc}}     & {\rm if} & u \ll 1 \\
     {\rm const.} & {\rm if} & u \gg 1
\end{array}
\right.,
\end{equation}   
it is only for $\alpha_{loc}=\alpha$ that
usual self-affine scaling \cite{FV} is recovered. 
This phenomenon has recently been elucidated 
by L{\'o}pez (see \cite{juanma} and references therein). 
In general it originates from an
additional correlation length, shorter 
than the system size, that enters as a relevant parameter in
scaling equations, destroying self-affinity.
In practical terms, it is important to observe that in the presence of 
anomalous roughening, if due attention is not paid (i.e., if scaling
relations are naively assumed to hold), one can measure
different correlation-time exponents depending on the type of experiment one performs.
Let us finally point out that the linear interface model,
at least in $d=1$, exhibits anomalous roughening \cite{arlim},
and therefore some
of the scaling anomalies we observe could be ascribed to effects 
of this nature. This is an issue that certainly deserves further study.

\section {Simulation results }

In this section we present numerical simulations of FES models. 
All three FES models studied here exhibit a critical point;
for large enough values of $\zeta$ the active site 
density (in the infinite-size limit)
has a nonzero stationary value. In order to study the critical point 
and the scaling behavior of the active state in
simulations of  finite systems, we must study the quasistationary
state that  describes the statistical properties of surviving trials.
The finite system size $L$, in fact, introduces a correlation 
length so that even above the critical point some initial 
configurations lead to an absorbing state.
In practice, we compute average properties  over a set
of $N_{samp}$ independent trials, 
each using a different 
initial configuration ($N_{samp}$ ranges from $10^3$ to $10^5$
depending on the lattice size). Quasistationary properties 
are calculated from averages restricted to surviving trials.
The active-site density exhibits the usual finite-size rounding 
in the neighborhood of the transition point; 
only in the limit $L\to\infty$ does
the transition become sharp. For this reason, finite-size scaling 
is a fundamental tool in the location of the critical point as well
as the calculation of critical exponents \cite{fss}. 

\subsection{The Manna FES model}

We performed simulations of the Manna fixed-energy
sandpile in the version in which the two particles
liberated when a site topples move independently to randomly chosen
nearest neighbors.
We studied lattices ranging from
$L = 32$ to  $1024$ sites on a side, using homogeneous, random
initial configurations as described in Sec. II.

\begin{figure}[bt]
\centerline{
        \epsfxsize=7.0cm
        \epsfbox{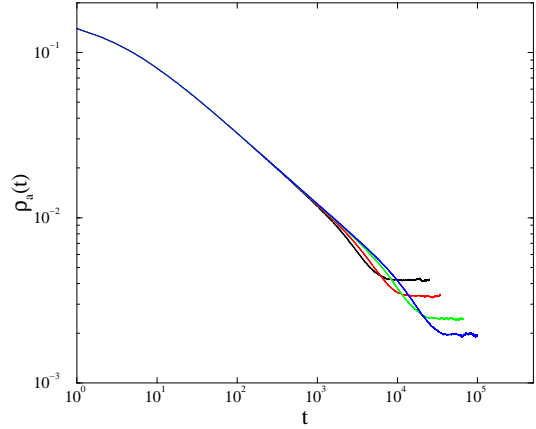}
        }
\caption{Manna FES: active-site density in surviving trials versus 
time at the critical point, $\zeta=0.71695$. From up to bottom,
system sizes $L=192, 256, 384, 512, 800$.}
\label{manna1}
\end{figure}

\begin{figure}[bt]
\centerline{
        \epsfxsize=7.0cm
        \epsfbox{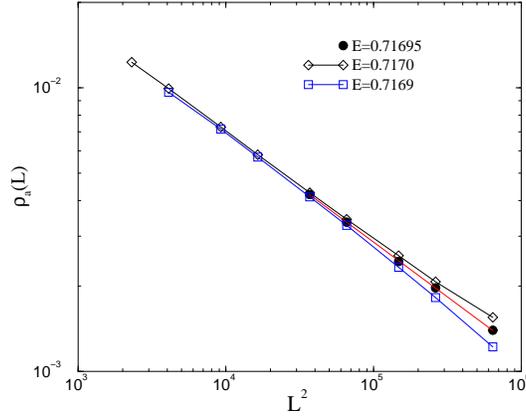}
        }
\caption{Stationary active-site density versus 
system size in the Manna FES. Sizes range from
$L=48$ to $L=800$.}
\label{manna2}
\end{figure}

After a transient whose duration depends on the system size $L$ and on
$\Delta\equiv\zeta-\zeta_c$, the surviving sample averages 
reach a steady value. In Fig.~(\ref{manna1}) we show how the density
of active sites approaches a mean stationary value 
$\overline{\rho_a}(\Delta,L)$. At a continuous transition
to an absorbing state, the order parameter ($\rho_a$ in this instance)
is expected to follow the finite-size scaling form

\begin{equation}
\overline{\rho_a} (\Delta,L) = 
L^{-\beta/\nu_{\perp}} {\cal R} 
(L^{1/\nu_{\perp}} \Delta) \;,
\label{actfss}
\end{equation}

\noindent where ${\cal R}$ is a scaling function 
with ${\cal R}(x) \sim x^{\beta}$ for large $x$, 
since for large enough $L>>\xi\sim \Delta^{-\nu_{\perp}}$ we must have 
$\overline{\rho_a}\sim \Delta^\beta$.  To locate $\zeta_c$ we study the 
stationary active-site density as a function of system size.
When $\Delta=0$ we have that $\overline{\rho_a} (0,L) 
\sim L^{-\beta/\nu_{\perp}}$; for $\Delta>0$, by contrast,
$\overline{\rho_a}$ approaches a stationary value, while for 
$\Delta<0$ it falls off as $L^{-d}$. Only at the critical point do we obtain 
a nontrivial power law, which allows us to locate the critical value 
$\zeta_c$. In Fig.~\ref{manna2} we observe power-law scaling 
for $\zeta = 0.71695$, but clearly not
for 0.7170 or 0.7169, allowing us to conclude that 
$\zeta_c = 0.71695(5)$. 
(Figures in parenthesis denote statistical uncertainties.)
The associated exponent ratio is $\beta/\nu_{\perp} = 0.78(2)$. 

\begin{figure}[bt]
\centerline{
        \epsfxsize=7.0cm
        \epsfbox{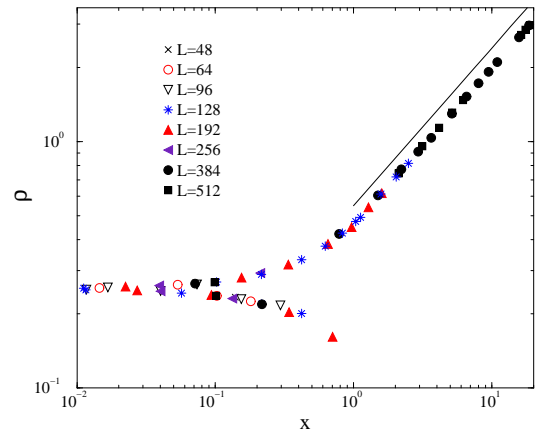}
        }
\caption{ Scaling plot of the stationary density 
$\rho \equiv L^{\beta/\nu_{\perp}} \overline{\rho_a}$ versus 
$x=L^{1/\nu_{\perp}} \Delta$ for various system sizes in the Manna FES. The slope 
of the straight line is $0.64$.} 
\label{manna3}
\end{figure}

\begin{figure}[bt]
\centerline{
        \epsfxsize=7.0cm
        \epsfbox{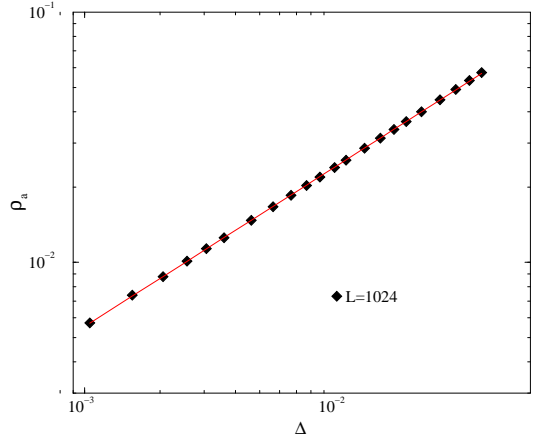}
        }
\caption{Stationary active-site density as a function of 
$\Delta=\zeta-\zeta_c$ for the Manna FES model with $\zeta_c=0.71695$.} 
\label{manna4}
\end{figure}

Next we consider the scaling behavior of the active-site
density away from the critical point.
The finite-size scaling form of Eq.~(\ref{actfss}) implies that a plot of 
$\rho  \equiv L^{\beta/\nu_{\perp}} \overline{\rho_a}$
versus $x \equiv L^{1/\nu_{\perp}} \Delta$ 
will show a data collapse
for systems of different sizes. 
In practice, we determine the horizontal and
vertical shifts (i.e., in a log-log plot of $\rho_a$ versus $\Delta$) 
required for a data collapse. In Fig.~\ref{manna3}, 
the best data collapse for $L \geq 48$ is obtained
with $\beta/\nu_{\perp} = 0.78(2)$ and $1/\nu_{\perp} = 1.22(2)$.
These values   correspond to an exponent $\beta=0.64(2)$. 
This is recovered also by a direct fitting of the scaling 
function ${\cal R}(x)$ for large $x$ (see Fig.~\ref{manna3}). 
A good estimate of $\beta$ can be also obtained by 
looking at the scaling of the stationary density with respect to 
$\Delta$ for the largest possible sizes $L$. In this case if $\Delta>0$
and $L>>\xi$ we have the scaling behavior $\overline{\rho_a}\sim \Delta^\beta$.
In Fig.~\ref{manna4}, we show the active site density as a function of 
$\Delta$ for $L=1024$. The resulting power-law behavior yields 
$\beta=0.64(1)$, where the error is dominated by the uncertainty in 
the critical point $\zeta_c$. 

\begin{figure}[bt]
\centerline{
        \epsfxsize=7.0cm
        \epsfbox{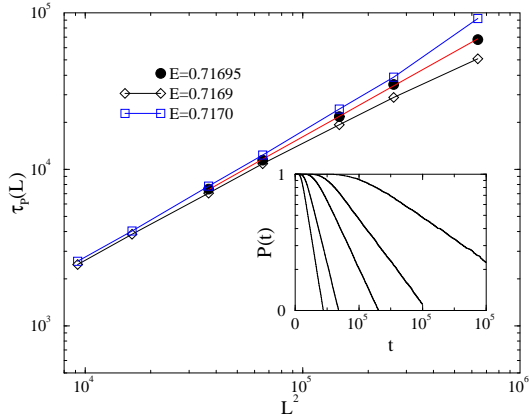}
        }
\caption{ Size dependence of $\tau_P$ close to the critical point of the Manna FES.
The inset shows the power law decay (on a linear-log scale) of the 
survival probability versus time at $\zeta_c=0.71695$ for 
sizes $L=192, 256, 384, 512, 800$ from
left to right.} 
\label{manna5}
\end{figure}

\begin{figure}[bt]
\centerline{
        \epsfxsize=7.0cm
        \epsfbox{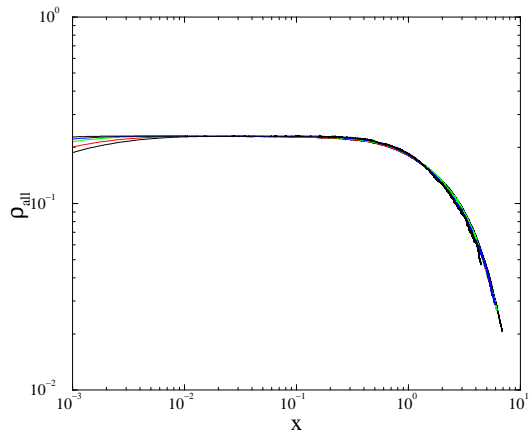}
        }
\caption{ Scaling plot of the scaled active-site density $\rho_{all}=\rho_{a,all}(t)t^\theta$,
in the Manna FES, averaged over all trials 
versus $x=tL^{-z}$ with $\theta=0.42(1)$ 
and $z=1.56(3)$. The system size ranges from $L=128$ to $L=800$.}
\label{manna6}
\end{figure}

To determine the dynamical exponent $z=\nu_{||}/\nu_{\perp}$ we study
the probability $P(t)$ that 
a trial has survived up to time $t$.
The latter appears to decay, for long times, 
as $P(t) \sim \exp(-t/\tau_P)$. At the critical point,
the characteristic decay time $\tau_P$ is a power-law function
of the only characteristic length in the system, the system
size $L$. Thus, we have $\tau_P(L)\sim L^z$ for $\Delta=0$. An estimate of 
$\tau_P(L)$ can be obtained by direct fitting of the exponential 
tail of $P(t)$, or by the time required for the survival probability 
to decay to one half. In Fig.~\ref{manna5} we report the behavior 
of $\tau(L)$ close to the critical point. Power-law behavior is
recovered at the critical point, yielding $z=1.57(4)$. (The error bar 
is again dominated by the uncertainty in the critical value 
$\zeta_c$.)  As a further consistency check we considered the 
density $\rho_{a, all}(t,L)$, that is, the active-site density 
averaged over all trials, including those that have reached the absorbing
state $\rho_a=0$. Assuming that the time dependence involves a 
single characteristic time that scales as $L^z$, we write at the 
critical point $\Delta=0$ 
\begin{equation}
\rho_{a, all}(t,L)= t^{-\theta}g(tL^{-z})
\end{equation}
where $g(x)$ is a constant for $x\ll1$ and decays faster than any power 
law for $x\gg 1$. A data collapse can be obtained by plotting 
$\rho_{all}=\rho_{a, all}(t,L)t^{\theta}$ versus $x=tL^{-z}$. 
The best data collapse is obtained with $\theta=0.42(1)$ and $z=1.56(3)$;
it is shown in Fig.~\ref{manna6}. This result confirms 
that the dynamical exponent is in the range  $z\simeq 1.55 -1.6$. 
An  exponent $\theta=0.42(1)$ is found also in the
decay of the active-site density $\rho_a(t)$ 
averaged only over the surviving trials
(see Fig.~\ref{manna1}). In simple absorbing-state transitions, 
the latter  exponent is consistent with the usual scaling 
relation $\theta = \beta/\nu_{||}$, obtained by assuming, for $\Delta=0$,
the simple 
scaling behavior $\rho_a(t)=L^{\beta/\nu_\perp}y(tL^z)$, with $y(x)=const$
for $x\to\infty$. In the Manna FES model, this simple scaling behavior 
is not observed, and the relaxation 
of the order parameter shows qualitatively
different scaling regimes. In particular, $\rho_a(t)$ exhibits a sharp drop 
(which seems to grow steeper with increasing $L$) just
before entering the final approach to $\overline{\rho_a}$
(see Fig.~\ref{manna1}). Accordingly, the exponent $\theta$ 
violates the usual scaling relation, and it is impossible to obtain 
a good data collapse with simple scaling forms. 
This is probably due 
to the introduction of an additional characteristic length that
defines the relaxation to the quasistationary state
(we are presently studying the possible relation between this effect
and anomalous roughening).
Moreover, it is not clear if the choice of initial conditions 
plays a role in this peculiar behavior.
A more detailed study of the relaxation to the 
stationary state is required in order to understand
the origin of these scaling 
anomalies, which appear in all the sandpile models analyzed in this paper,
as well as in the one-dimensional Manna FES \cite{man1d}.

\begin{figure}[bt]
\centerline{
        \epsfxsize=7.0cm
        \epsfbox{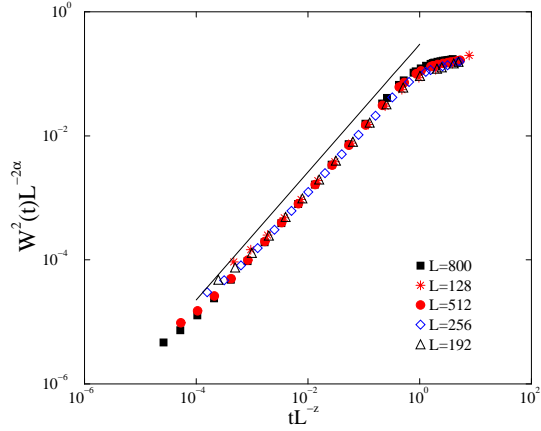}
        }
\caption{ Data collapse analysis at $\zeta_c=0.71695$ for the interface width $W(t,L)$ of
$H(i,T)$, defined as the total number of toppling at time $t$ 
for each site $i$, in the Manna FES. The exponents used are $\alpha=0.81(2)$ and $z=1.58(3)$.}
\label{manna7}
\end{figure}

The interface mapping described in Sec. IV 
prompted us to study the dynamics of
the mean width $W(t,L)$ [see Eq.(\ref{w2})].
We studied the
evolution of the width at $\zeta_c$, in systems of size $L= 128$ to 800.
Unfortunately, we were not able to reach the complete saturation regime of the 
roughness, which would afford an independent estimate of the exponent $\alpha$.
This is due to the exponential decay of the survival probability at
very large times. 
As shown in Fig.~\ref{manna7}, we obtain a good collapse 
using the values $\alpha = 0.80(3)$ and $z = 1.57(2)$.  
Following Eq.~(\ref{w2scal}), the short-time behavior of $W(t,L)$ gives 
an exponent $\beta_W=0.51(1)$. This exponent, however, shows a systematic
increase with the system size $L$. In particular, for large sizes ($L\geq 512$)
it seems that a simple power-law regime is not adequate to represent the 
temporal behavior of the interface width. 
Note also that the scaling relation $\theta + \beta_W = 1$,
satisfied to within uncertainty for the other models considered,
is violated in the Manna case: $\theta + \beta_W = 0.93(2)$.
It appears that some of the anomalies affecting the 
temporal scaling of surviving
trials could be influencing the estimates of the roughness exponents.
Also in this case, further studies, for example of the local roughness,
are needed for a direct comparison with other interface growth models. 

In summary, numerical results show clear evidence of the critical 
behavior usually observed in absorbing phase transitions. Critical exponents 
and a discussion about universality classes will be provided in the next 
section. Finally, we note that the Manna sandpile does not 
exhibit the strong nonergodic effects reported below for the BTW model.

\subsection{The BTW FES model}

In Ref. \cite{dvz} preliminary
results on the two-dimensional BTW model were reported; here we
present a more detailed study, including considerably larger lattices.  
To study stationary properties,
we performed, for each system size $L$ = 20, 40,...1280, and energy density $\zeta$,
$N_{samp}$ independent trials (ranging from $5 \times 10^4$ for $L=20$ to
1600 for $L=1280$), each extending up to a maximum time $t_{max}$.
The latter, which ranged from 800 for $L=20$ to $3 \times 10^5$ for
$L=1280$, was sufficient to probe the stationary state. 
An overall idea of the dependence of the active-site density on $\zeta$
can be gotten from  Fig.~\ref{mffig1}, which compares simulation results
with the pair approximation derived in the Appendix.

\begin{figure}[bt]
\centerline{
        \epsfxsize=7.0cm
        \epsfbox{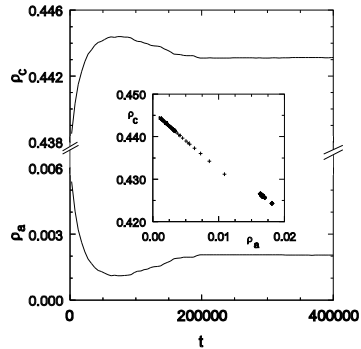}
        }
\caption{ Relaxation of the active-site density $\rho_a$ (lower graph) 
and the critical-site
density $\rho_c$ (upper graph) in the BTW FES 
at the critical point
($\zeta = 2.125$, $L=1280$).  Inset: scatter plot 
of $\rho_c$ versus $\rho_a$; $\times$:
$\zeta=\zeta_c$, $L=1280$; $+$: $\zeta=\zeta_c$, $L=640$; 
diamonds: $\zeta = 2.13$,
$L=320$.}
\label{btw1}
\end{figure}

The simulations reported in Ref. \cite{dvz}, which extended to systems
of linear dimension $L=160$, permitted us to conclude that
$\zeta_c = 2.1250(5)$ \cite{exact}.
 We first discuss the results of simulations
performed at $\zeta_c$.  Figure~\ref{btw1} shows the relaxation of the active-
and critical-site densities at $\zeta_c$; note the non-monotonic
approach to the limiting values.  The inset shows that there is
a deterministic, linear relation between the two densities during 
the relaxation process:
for $\zeta = \zeta_c$, a least-squares fit yields
$\rho_c  = \rho_{c,cr} - C \rho_a$, where $C=1.368$ 
and $\rho_{c,cr} = 0.4459$ is the
critical site density at $\zeta_c$ in the limit 
$L \rightarrow \infty$ (for which
$\rho_a$ naturally falls to zero).  
We note that this relation is independent of
system size $L$ and of sample-to-sample variations (for the same $L$);
all that changes is the portion of the line filled in by the data.
For off-critical values of the energy density, the
active- and critical-site densities follow
a different linear trend \cite{annote}.

\begin{figure}[bt]
\centerline{
        \epsfxsize=7.0cm
        \epsfbox{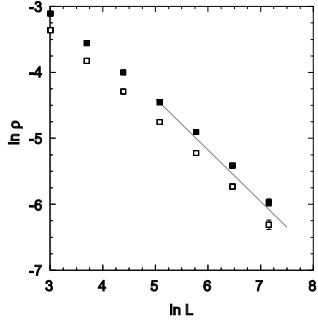}
        }
\caption{Stationary active-site density (open squares) 
and excess critical-site 
density (filled squares)  versus system size in the 
BTW FES at $\zeta_c$.}
\label{btw2}
\end{figure}

In Fig.~\ref{btw2} we plot $\overline{\rho_a} (\zeta_c,L) $ 
and the excess critical-site density 
$|\overline{\rho_c} (\zeta_c,L) - \zeta_{c,cr}|$
(overbars denote mean stationary values),
versus $L$ on log scales, anticipating that these  decay 
$ \sim L^{-\beta/\nu_{\perp}}$.
The apparent power-law behavior for small $L$ 
is followed, for larger $L$, by an
approach to a larger exponent.  For $L \geq 320$ we obtain
the estimates of  $\beta/\nu_{\perp} = 0.78(3)$ and 0.77(2) from 
the active-site and critical-site density, respectively, but it is clear
that the slope of this plot has not stabilized even for $L=1280$.

\begin{figure}[bt]
\centerline{
        \epsfxsize=7.0cm
        \epsfbox{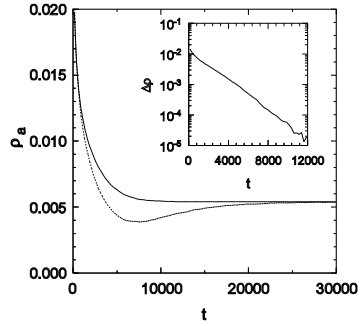}
        }
\caption{Relaxation of the active-site density in the 
BTW FES at $\zeta_c$ ($L=320$).
Dashed line: unrestricted sample; solid line: 
sample restricted to runs surviving to $t_{max} = 10^5$.  
The inset is a semilog plot of $\rho_a (t)$ for the restricted sample.}
\label{btw3}
\end{figure}

\begin{figure}[bt]
\centerline{
        \epsfxsize=7.0cm
        \epsfbox{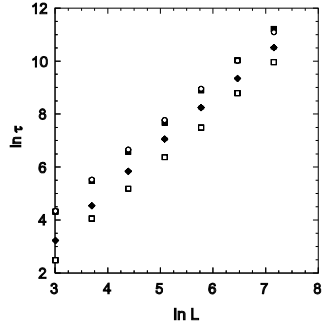}
        }
\caption{Relaxation times versus system size in the 
BTW FES at 
$\zeta_c$. Open squares: $\tau_a$; 
filled squares: $\tau_m$; diamonds: $\tau_P$; circles:
$\tau_{\overline{P}}$.}
\label{btw4}
\end{figure}

Next we consider the relaxation time at $\zeta_c$.
There are two independent quantities whose relaxation is readily
monitored: the survival probability $P(t)$ and the active-site density
$\rho_a (t)$. (Given the strict linear relationship between $\rho_a$ and $\rho_c$,
we cannot treat the latter as an independent dynamical variable; not
surprisingly, the two yield essentially the same relaxation times.)  
We studied four different relaxation times; the first two are associated with
the survival probability $P(t)$.  This quantity decays 
slowly at first, then enters a
regime of roughly exponential decay, after which it attains a nearly
constant value $P_P$.  (While $P(t)$ appears to decay very slowly
after attaining $P_P$, the relaxation times we study here are 
for the approach to $P_P$.)
We define $\tau_P$ as the relaxation time associated with the exponential-decay
regime; another relaxation time, $\tau_{\overline{P}}$, 
is defined as the time at which
$P(t)$ equals $(1+P_P)/2$, midway to its quasi-stationary value.
As we have seen, $\rho_a (t)$ exhibits a non-monotonic approach to its
stationary value, and does not exhibit a clear exponential regime.
Taking advantage of the non-monotonicity, we define $\tau_{m}$
as the time at which $\rho_a$ takes its minimum value.
Finally, we noted that {\it restricting} the sample to trials
that survive up to $t_{max}$ results in a monotonic, exponential approach to
$\overline{\rho_a}$ (see Fig.~\ref{btw3} ).   
A fit to the linear portion of a semi-log plot of the excess density 
$\rho_a (t) - \overline{\rho_a}$ yields $\tau_a$.
Relaxation times in a critical system are expected to diverge with system size 
as $\tau (\zeta_c,L) \sim L^{\nu_{||}/\nu_{\perp}}$.
The data for all four relaxation times, plotted in 
Fig.~\ref{btw4}, are consistent with a power law, but
due to fluctuations, linear fits to the data (for $L \geq 160$) 
yield exponent ratios ranging from $\nu_{||}/\nu_{\perp} = 1.59$ to 1.74.  
Since the four data sets do seem to follow a common trend, and since there is no reason 
to expect different relaxation times to be governed by different exponents, we 
define $\overline{\tau}(L)$ as the geometric mean of all 
four relaxation times.  The behavior of 
$\overline{\tau}(L)$ is quite regular; linear fits to 
the data for $L \geq 80$, 160, and 320 yield 
$\nu_{||}/\nu_{\perp} = 1.671$, 1.668 and 1.657, 
respectively, leading to an estimate of 1.665(20) for this ratio.

\begin{figure}[bt]
\centerline{
        \epsfxsize=7.0cm
        \epsfbox{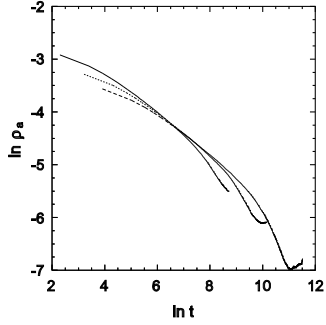}
        }
\caption{ Initial decay of the active-site 
density in the BTW FES at 
$\zeta_c$. Solid line: $L=320$; dotted line: $L=640$; dashed line: $L=1280$.}
\label{btw5}
\end{figure}

Another manifestation of scaling is the short-time decay of the
order-parameter density in a critical system, starting from a
spatially homogeneous initial configuration \cite{short}.
In Fig.~\ref{btw5} we show the active-site density for short-times.
The data exhibit an imperfect collapse, and there is no clear-cut power-law
regime.  The roughly linear region for $L=1280$
yields a decay exponent $\theta \simeq 0.41$.

\begin{figure}[bt]
\centerline{
        \epsfxsize=7.0cm
        \epsfbox{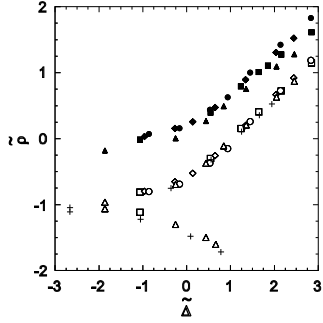}
        }
\caption{Scaled order parameter $\tilde{\rho}$ versus 
scaled distance from criticality
$\tilde{\Delta}$ in the BTW FES.  
Symbols for the scaled active-site density:
$+$: $L=40$; $\triangle$: $L=80$; $\Box$: $L=160$; 
$\Diamond$: $L=320$; $\circ$: $L=640$.
The filled symbols represent the scaled excess 
critical-site density $\tilde{\rho_c}$ for
$L=80$, 160, 320, and 640.} 
\label{btw6}
\end{figure}

Next we consider the scaling behavior of the active- and critical-site
densities away from the critical point.  
We analyze these data using the finite-size scaling form 
of Eq. (\ref{actfss}), which implies that a plot of 
$\tilde{\rho} \equiv L^{\beta/\nu_{\perp}} \overline{\rho_a}$
versus $\tilde{\Delta} \equiv L^{1/\nu_{\perp}} \Delta$ will 
show a data collapse for systems of different sizes.  
The data analysis is as described above for the Manna FES.
The best data collapse (see Fig.~\ref{btw6}) for $L \geq 80$ is obtained
with $\beta/\nu_{\perp} = 0.75(2)$ and $1/\nu_{\perp} = 1.15(2)$.
(This value of $\beta/\nu_{\perp}$ is slightly smaller than the 
value obtained above
from the scaling of $\rho_a$ at $\zeta_c$; note however that the latter value,
0.78(3), is based on systems with $L \geq 320$.)
>From this finite-size scaling analysis we therefore obtain the 
values $\nu_{\perp} = 0.87(2)$ and $\beta = 0.65(2)$.  One again, though, it is 
important to check for size dependence of the exponent estimates.  
Fitting the linear portion of the $\rho_a$ data in the scaling
plot, we obtain $\beta = 0.62$, 0.63, 0.66 and 0.69 for $L = 80$, 160, 320 
and 640, respectively.

We can apply a similar analysis to the density of critical sites,
but here we must
isolate the {\it singular part} of $\rho_c$ from an analytic background.  
The latter appears
because for $\zeta < \zeta_c$, $\rho_c$ increases smoothly
with $\zeta$.  Above $\zeta_c$, $\rho_c$ decreases linearly with 
$\rho_a \sim \Delta^{\beta}$, so we expect the singular part 
$\rho_{c,sing} = A\Delta^{\beta}$ for $\Delta > 0$, with $A<0$.  
The simplest reasonable 
form for the nonsingular background 
is $\rho_{c,reg} = \rho_{c,cr} + B\Delta$, where
$\rho_{c,cr} = 0.4459$ is the 
$L \rightarrow \infty$ critical value as noted above.
We expect the singular part of $\rho_c$ 
to follow the same finite-size scaling form as
the active-site density.  This implies that

\begin{equation}
\rho_c^* (\tilde{\Delta}, L) 
\equiv L^{\beta/\nu_{\perp}} (\rho_c - \rho_{c,cr})
                    = -C{\cal R}(\tilde{\Delta}) 
                       + BL^{(\beta-1)/\nu_{\perp}} \tilde{\Delta} \;.
\label{rhocsc}
\end{equation}
Thus the singular contributions cancel in $\rho_c^* (L) - \rho_c^*(L')$.
Using the values for $\nu_{\perp}$ and $\beta/\nu_{\perp}$ 
found in the scaling analysis
of $\rho_a$, we study $\rho_c^* (L) - \rho_c^*(L')$ 
for all pairs of system sizes in the range
$L = 80,...,640$, and obtain $B=0.71(2)$.
We can then construct a scaling plot of the singular part,
$\tilde{\rho}_{c,sing} \equiv L^{\beta/\nu_{\perp}} 
|\rho_c - \rho_{c,cr} - B \Delta|$,
which shows a fair data collapse (see Fig.~\ref{btw6}), 
but with much more scatter than  for
$\rho_a$, presumably because of the uncertainties 
involved in isolating the singular contribution.
As in the case of the active-site density, the $\beta$ estimates we
obtain from the $\rho_{c,sing}$ data increase with $L$.
Here we find $\beta = 0.65$, 0.65, 0.67 and 0.70 for $L = 80$, 160, 320 and
640, respectively.  
We conclude that $\beta \stackrel > \sim  0.7$.  Studies of larger 
lattices will be required to refine this estimate.

\begin{figure}[bt]
\centerline{
        \epsfxsize=7.0cm
        \epsfbox{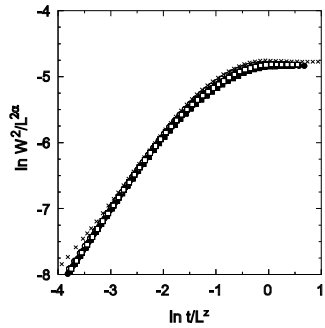}
        }
\caption{Scaling plot of the mean-square interface width $W^2(t,L)$ in the 
BTW FES.  $\times$: $L=40$; $\circ$: $L=80$; 
$\bullet$: $L=160$; $\Box$: 
$L=320$; filled squares: $L=640$.}
\label{btw7}
\end{figure}

\begin{figure}[bt]
\centerline{
        \epsfxsize=7.0cm
        \epsfbox{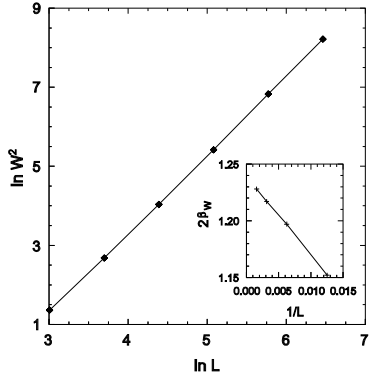}
        }
\caption{Main graph: saturation value of the mean-square 
interface width $W^2 $
versus system size $L$ in the
BTW FES at $\zeta_c$.  Inset: apparent value of
the growth exponent $\beta_W$ plotted versus $1/L$.}
\label{btw8}
\end{figure}

We studied the evolution of the interface width $W(t,L)$ as defined in Eq~(\ref{w2}),
at $\zeta_c$, in systems of size $L= 20$ to 640,
with sample sizes ranging from $5 \times 10^4$ for $L=20$ 
to $10^3$ for $L=640$.
As shown in Fig.~\ref{btw7}, we obtain a good collapse for $L \geq 40$
using the values $\alpha = 1.01(1)$ and $z = 1.63(2)$.  
The exponent $\alpha $
can be found directly from the data for the saturation 
value of $W^2$ shown in Fig.~\ref{btw8}.
Fitting the short-time (power-law) data for $W^2$ yields an estimate for 
the growth exponent $\beta_W$, which increases systematically with $L$,
as shown in the inset of Fig.~\ref{btw8}.
Extrapolating to infinite $L$ we obtain 
$\beta_W = 0.62$, in agreement with the scaling
relation $\beta_W = \alpha/z$ .  Note also that the value of $z$ 
describing the interface growth crossover time is consistent, as one would expect, with that 
for $\nu_{||}/\nu_{\perp}$, derived from a study of relaxation times.

\begin{figure}[bt]
\centerline{
        \epsfxsize=7.0cm
        \epsfbox{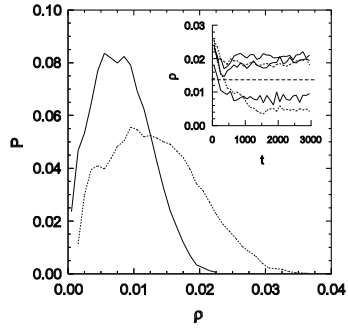}
        }
\caption{Main graph: histograms for the stationary mean active-site 
density in a given trial
in the BTW FES at $\zeta_c$.  
Dashed line: $L=80$; solid line: $L=160$.
The inset shows the evolution of the active-site 
density in five different trials ($L=80$); the
dashed line represents the stationary mean value 
averaged over a large number of trials.}
\label{btw9}
\end{figure}

The size dependence of the critical exponents 
could be an indication of the failure of the simple scaling
hypothesis \cite{stella}.
A further anomalous aspect of the BTW FES is
{\it nonergodicity}:
in a particular trial, properties such as $\rho_a$ typically differ from
the mean value computed over a large number of trials.
This means that time averages are different from averages over 
initial configurations, where the latter play the role of 
``ensemble averages''. It is worth remarking that this nonergodicity
is consistent with the existence of toppling invariants \cite{dhar}. 
In Fig.~\ref{btw9}, for example, we show the evolution of $\rho_a$ for
five different initial configurations (ICs) in a 
system with $L=80$, at $\zeta_c$.
Each IC appears to yield a
particular active-site density; fluctuations about this value are
quite restricted, and typically do not embrace the mean over ICs.
Fig.~\ref{btw9} also shows histograms of the stationary 
mean active-site density (for a given IC), in samples of 10$^4$ ICs, 
for $L=80$ and 160; the distribution has a 
single, well-defined maximum, and narrows with increasing $L$.
The data indicate, however, that the probability distribution for 
$\rho_a/\overline{\rho_a}$ (i.e., the order parameter normalized
to its mean value), does not become sharp as $L \rightarrow \infty$,
as it would, for example, in directed percolation.

\begin{figure}[bt]
\centerline{
        \epsfxsize=7.0cm
        \epsfbox{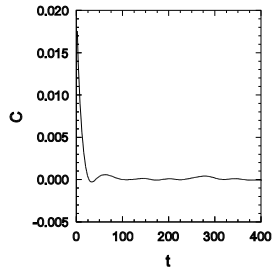}
        }
\caption{Autocorrelation function for the number of active sites in the 
BTW FES at $\zeta_c$, ($L=80$) averaged over 2000 trials.}
\label{btw10}
\end{figure}

Further evidence of nonergodicity is found in the activity autocorrelation
function, defined as

\begin{equation}
C(t) \equiv \frac{\langle N_A(t_0+t)N_A (t_0) \rangle} 
{\langle N_A (t_0) \rangle^2} -1 ,
\label{ctdef}
\end{equation}
where $N_A(t)$ is the number of active sites at time $t$, and 
$\langle ... \rangle$ stands for an average over times $t_0$ 
{\it in the stationary state} for a given IC, as well as
an average over different ICs.  The autocorrelation function
for the critical BTW FES ($L=80$, average over 2000 ICs and $10^4$ time
units), shown in Fig.~\ref{btw10}, exhibits surprisingly little structure.  
After decaying rapidly to a
minimum value at around $t=34$, and increasing to a weak local maximum near
$t=62$, $C(t)$ seems to fluctuate randomly about zero.  The relaxation occurs
on a time scale over an order of magnitude smaller than for $\rho_a$ or the
survival probability (the relaxation times 
$\tau_{m}$ and $\tau_{\overline{P}}$
$\approx 800$ for this system size).

\begin{figure}[bt]
\centerline{
        \epsfxsize=7.0cm
        \epsfbox{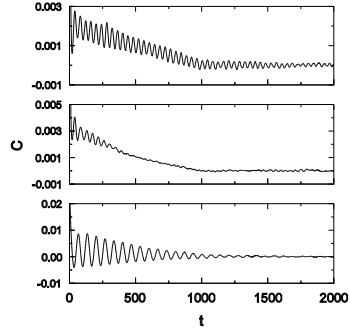}
        }
\caption{Autocorrelation function for the number of active sites in the 
BTW FES at $\zeta_c$, ($L=80$) in three different trials.}
\label{btw11}
\end{figure}

\begin{figure}[bt]
\centerline{
        \epsfxsize=7.0cm
        \epsfbox{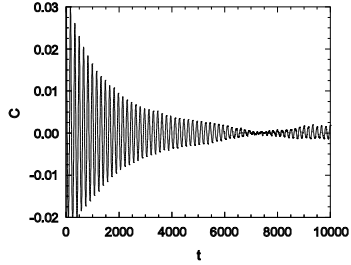}
        }
\caption{Autocorrelation function for the number of active sites in the 
BTW FES at $\zeta_c$, ($L=80$) in a long trial.}
\label{btw12}
\end{figure}

The reason for this anomalously rapid decay becomes clear when we examine
the autocorrelation function in individual trials ($C(t)$ defined 
as in Eq. (\ref{ctdef})
but {\it without} averaging over ICs).  Figures~\ref{btw11} and \ref{btw12}   
show some typical 
results for $L=80$.  (Here, to get good statistics, we have averaged
over $5 \times 10^5$ to $10^6$ time units in the stationary state.) 
The correlation function in a single trial shows 
shows considerable structure, including damped oscillations 
(and in some cases, revivals),
which may be superimposed on a more-or-less linear decay.
The period (in the range 35 - 70 for $L=80$) and other
features vary from one IC to another.  (Changing the seed for the
random choice of toppling sites changes $C(t)$ only slightly,
if we maintain the same IC \cite{unote}.)  Evidently, $C(t)$ 
decays rapidly to zero when we average over initial conditions
because of dephasing amongst oscillatory signals with varied frequencies.
Interestingly, the interface width $W(t,L)$ shows much 
less dependence on the IC than does the 
active-site density or its autocorrelation.

In summary, the BTW fixed-energy sandpile shows 
signs of the kind of scaling found at simpler 
absorbing-state phase transitions, 
but at the same time exhibits dramatic nonergodic effects.  
We note unusually strong finite-size effects, which
prevent us from determining certain critical exponents precisely.
Whether this is a simple finite-size effect or 
a signature of multiscaling cannot be ascertained definitively 
with the present data.

\subsection{The Shuffling FES  model}

The shuffling model\cite{zhang2} has a continuously variable control parameter,
since each site has a (non-negative) real-valued energy. Thus we are no longer
constrained to vary the energy density $\zeta$ in increments of
$1/L^2$ as we are in discrete models (e.g., the Manna and BTW FES), 
where the single grain is the smallest
energy unit. In 
the shuffling FES, all sites whose energy exceeds the threshold 
$z_{th}=2$ are considered active. In addition, sites that have received energy
from a toppling nearest neighbor can become active if $z_i< z_{th}$ 
with  a probability $p_i=z_i/z_{th}$. 
This enlarges considerably the choice of possible initial configurations. 
In particular, after we have  distributed randomly the total amount of energy
among the lattice sites, 
we extract for each site a random number $\eta_i$ and we declare 
active all sites for which $\eta_i\le z_i/z_{th}$. (Obviously, sites 
with $z_i\ge z_{th}$ are active with probability one.) 
Unlike discrete models, we have the option of generating ``flat'' 
initial conditions,
in which all sites have the same energy.
While stationary properties are not affected by the choice 
of noisy versus flat initial 
configurations, we do note differences in the short-time behavior.

Another peculiar characteristic of the shuffling model 
is the strong non-Abelian 
character of its dynamics. We implemented the dynamics 
of the model with parallel updating as in the original definition of 
Ref.~\cite{zhang2}. However, this form of the dynamics contains 
some non-local effects as described in Sec. II, and does not ensure
that parallel and  sequential updating generate the same critical behavior.
Simulations with sequential updating are in progress.

\begin{figure}[bt]
\centerline{
        \epsfxsize=7.0cm
        \epsfbox{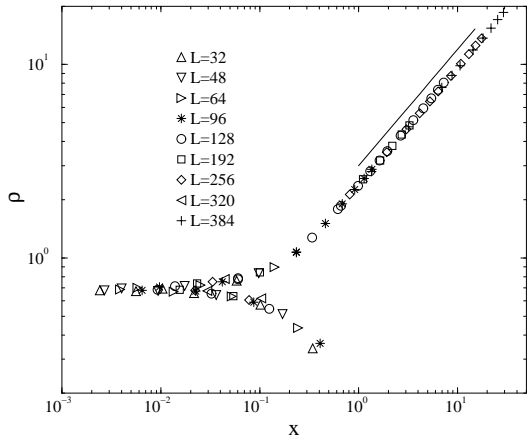}
        }
\caption{Scaling plot of the stationary active-site density 
$\rho \equiv L^{\beta/\nu_{\perp}} \overline{\rho_a}$ versus 
$x=L^{1/\nu_{\perp}} \Delta$ for various system sizes 
in the shuffling model. Here 
$\beta/\nu_{\perp}=0.76$ and $1/\nu_{\perp}=1.266$. The slope 
of the straight line is $0.60$.}
\label{shuffle1}
\end{figure}

Simulations of the shuffling model require many 
calls to the random number generator,
and so are extremely time-consuming.
Here we present simulations with flat initial conditions and 
sizes ranging from $L=32$ to $L=384$.
By analyzing the $L$-dependence of $\overline{\rho_a}(\Delta,L)$
we find the critical point $\zeta_c=0.20427(5)$. 
When $\zeta=\zeta_c$ the stationary density has a power-law behavior
$\overline{\rho_a}(0,L)\sim L^{\beta/\nu_\perp}$ that
yields $\beta/\nu_\perp=0.76(3)$. This result is confirmed by the scaling plot
of Fig.~\ref{shuffle1}, which, following Eq.~(\ref{actfss}) shows    
$\rho  \equiv L^{\beta/\nu_{\perp}} \overline{\rho_a}$
versus $x \equiv L^{1/\nu_{\perp}} \Delta$, with $\beta/\nu_{\perp}=0.76$
and $1/\nu_{\perp}=1.266$. This gives an exponent $\beta=0.60$, as confirmed 
by the straight slope of the upper branch of the scaling plot. 
An independent measurement of the stationary density versus $\Delta$ for
the largest size used ($L=384$) gives the estimate $\beta=0.60(2)$, where 
the error bar is due mainly to the uncertainty in $\zeta_c$.

\begin{figure}[bt]
\centerline{
        \epsfxsize=7.0cm
        \epsfbox{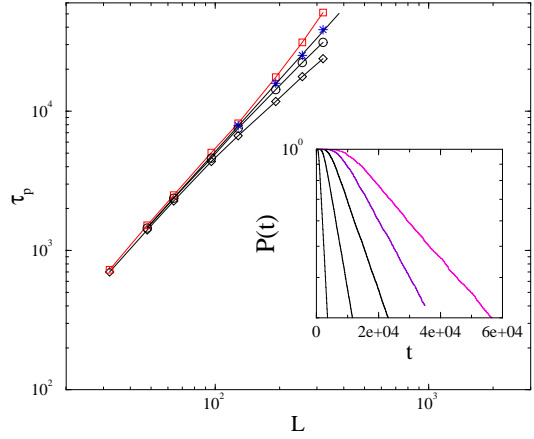}
        }
\caption{Size dependence of $\tau_P$ close to the critical 
point of the shuffling FES.
$\diamond$: $\zeta=0.2420$; $\circ$: $\zeta=0.2425$; 
$\ast$: $\zeta=0.2427$; $\Box$: 
$\zeta=0.2430$. 
The inset shows the power-law decay (on a linear-log scale) of the 
survival probability versus time at $\zeta_c=0.20427$ for 
sizes $L=128, 192, 256, 320$ from
left to right.}
\label{shuffle2}
\end{figure}

\begin{figure}[bt]
\centerline{
        \epsfxsize=7.0cm
        \epsfbox{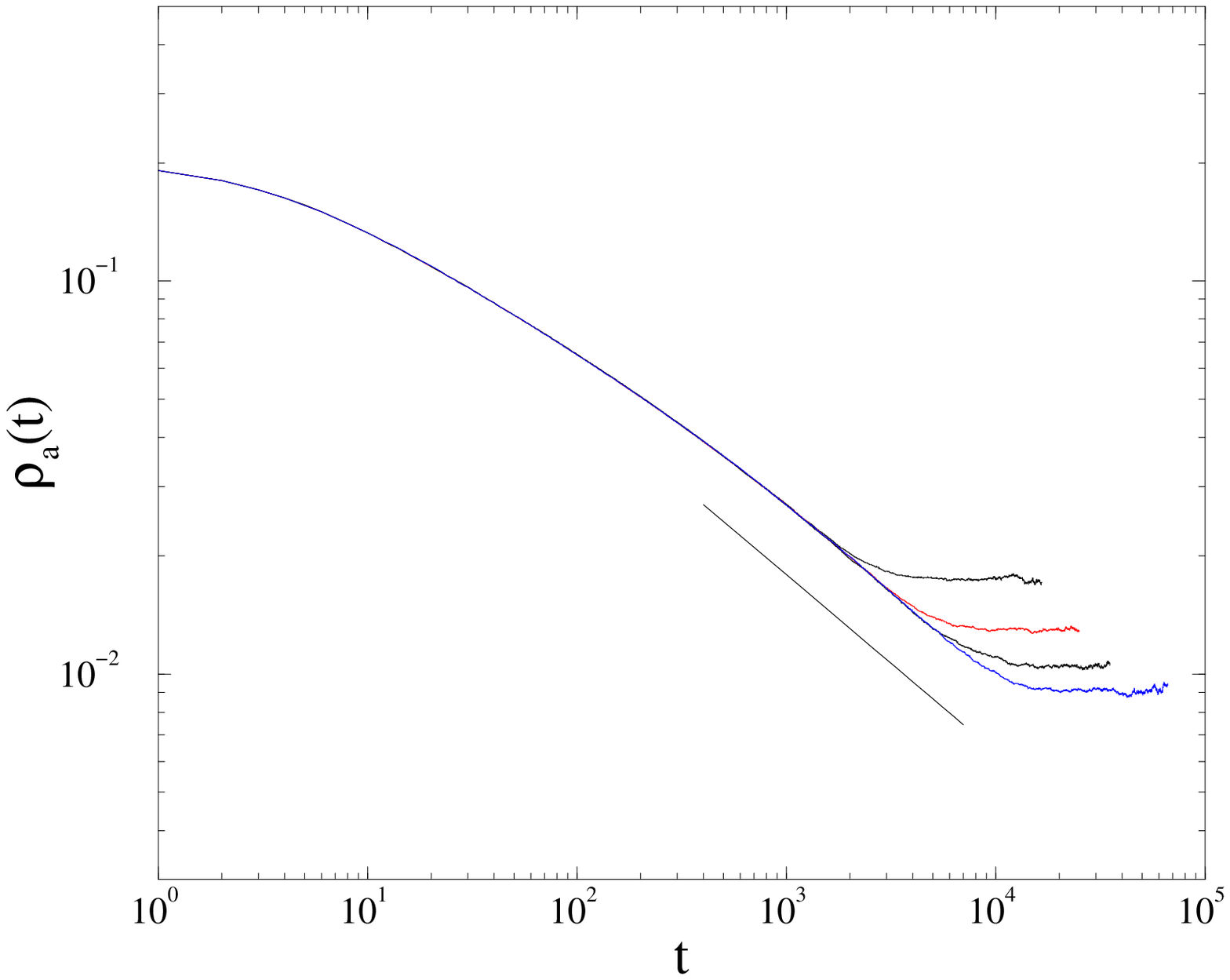}
        }
\caption{ Shuffling FES: active-site density in surviving trials versus 
time at the critical point $\zeta=0.20427$. From up to bottom 
system sizes $L=128, 192, 256, 320$. The 
straight line has  slope $\theta=0.45$.   }
\label{shuffle3}
\end{figure}

We performed a scaling analysis of the temporal behavior by studying
the decay of the survival probability $P(t)\sim\exp(-t/\tau_P)$.
At the critical point the $L$-dependence of the characteristic time 
assumes the power-law behavior $\tau_P\sim L^z$ with $z=1.71(5)$ 
(see Fig.~\ref{shuffle2}).  
However, it is worth noting that the scaling behavior with $L$ shows a 
systematic curvature from the smallest to the largest sizes, both below and 
above the critical point. This could be a signal that the system has not 
yet reached its asymptotic temporal behavior for the sizes 
considered ($L \leq 320$). 
That the relaxation could be affected by strong 
finite-size effects is confirmed by the temporal scaling of 
$\rho_a(t,L)$. In Fig.~\ref{shuffle3} we observe that the active-site density 
decay does not follow a definite power law before 
reaching the stationary state. This makes impossible an accurate 
determination 
of the exponent $\theta$ ($\approx 0.46$), which is also reflected in the 
absence of a clear data collapse for the temporal scaling functions. 

The roughness analysis is affected by several numerical problems. The 
short average lifetime of trials at finite size makes it impossible to 
reach the width-saturation regime. This effect is even more pronounced than 
in the Manna case. It is therefore impossible to apply a 
data-collapse analysis,
nor is a direct measurement, that would yield $\alpha$, feasible. 
The short-time 
behavior of the roughness (see Eq.~(\ref{w2})) is governed by the exponent 
$\beta_W\simeq 0.57$. Applying the scaling relation shown in Sec. IV,
and using the dynamical exponent obtained previously,
we have $\alpha\simeq 0.96$.
However, in this case the short-time behavior of the roughness 
appears to have a size dependence, probably due to the lack 
of complete convergence to the asymptotic scaling behavior, and the 
numerical values provided here could contain systematic errors that are
difficult to estimate.
 
In summary, the numerical results for the shuffling FES model show also 
the signature of a continuous phase transition from an absorbing
to an active phase. 
The stationary properties  of the model show well defined 
scaling behavior at the system sizes considered in the present study. 
The dynamic scaling properties, by contrast, show anomalies 
and transient effects that could indicate that the system has not yet 
attained its asymptotic behavior for $L \leq 384$.

\section{Discussion and open questions} 

\subsection{Universality classes and critical exponents}

Simulations of sandpile models have mainly been performed in 
the slow driving regime. It is then natural to  
compare the critical exponents measured in the fixed-energy framework 
(see Tab.~\ref{exps}) with those observed in driven simulations. 
In driven sandpiles, critical behavior is characterized 
by the scaling of the number of topplings $s$ and the duration $t$ 
following the addition of an energy grain\cite{btw},
i.e., an  avalanche. The 
probability distributions of these variables are usually described 
with the finite-size scaling forms 
\begin{equation}
P(s)=s^{-\tau_s}{\cal G}(s/s_c)
\end{equation}
\begin{equation}
P(t)=t^{-\tau_t}{\cal H}(t/t_c)
\end{equation}
where $s_c\sim L^D$ and $t_c\sim L^z$ are the characteristic avalanche 
size and time, respectively. Applying the fundamental result (due to 
conservation), $<s>\sim L^2$\cite{dhar,vz,tb88}, we can write 
the scaling relations 
$\tau_s=2-2/D$ and $\tau_t=1+(D-2)/z$.
Recently, these  simple scaling forms have been 
questioned in the case of the BTW model\cite{stella}. 
An accurate moment analysis seems 
to show multiscaling, so that scaling relations obtained from the 
above finite-size scaling forms do not apply.

While critical exponents governing the  
deviations from criticality in FES do not have any counterpart in 
the driven case, which is posed by definition at the critical point, 
the exponents describing the critical point, including $z$ and 
the fractal dimension $D$,
can be compared directly. In FES simulations $D$ can be calculated by noting
that the scaling of an avalanche due to a point seed scales as the total 
variation of the field $H(i,t)$, which represents the total 
number of topplings. 
Since the roughness scales with exponent $\alpha$, we readily obtain that 
$D=d+\alpha$\cite{pacz,lau}. 

For the Manna model, our simulations yield $D=2.80(3)$ and $z=1.57(4)$, 
which should be compared with the most recent analyses of driven sandpiles,
which yield $D=2.76(2)$ and $z=1.56(2)$\cite{csvz,lubeck2,cvz,romu}. 
By using scaling relations 
we obtain $\tau_s\simeq 1.29$ and $\tau_t\simeq 1.51$, again in very 
good agreement with the values obtained in the driven case.
For the shuffling model we can compare our results $z=1.71$ and $D=2.96$ 
with the simulations of Maslov and Zhang\cite{zhang2},
which give $z=1.73(5)$ and $D=2.92(5)$. In 
this case we also see a very good agreement between independent measurements.

More subtle is the case of the BTW model. Here different 
simulations of the driven sandpile give rather scattered results. 
A very recent analysis suggesting multiscaling in 
the (driven) BTW sandpile indicates that  
neither $D$ nor $z$ are clearly defined\cite{stella}. 
In particular, the effective value of $D$ increases as one studies
higher moments, and saturates at $D\simeq 3.0$. This is indeed the 
result we recover from our analysis ($D=3.01(1)$). The possibility of 
multiscaling is supported by the scaling anomalies and the 
lack of self-averaging we detected in our simulations of the BTW FES.

We shall attempt, on the basis of our numerical results, 
to assign the various fixed-energy
sandpiles studied to universality classes.
This a particularly vexing problem, that has eluded 
ten years of theoretical and numerical efforts.
Soon after the introduction of  sandpile models with modified
dynamical rules, there were
many quests for the precise identification of universality classes. 
In particular BTW and Manna models, which are prototypes for 
deterministic and stochastic models, respectively, have been 
the objects of a longstanding quarrel over their supposed 
universality classes\cite{manna,grasma,lubeck1,ben,csvz,lubeck2,cvz}. 
The first numerical attempts showed 
very similar exponents for the avalanche distributions\cite{manna,grasma}, 
but the results were afflicted by severe finite-size errors due to the 
limited sizes attainable using the CPU power available at that time. 
These results were later questioned by Ben-Hur and Biham\cite{ben},
who analyzed the scaling of conditional expectation values 
of various quantities related to avalanches. These results were, 
however, biased by the unexpected singular behavior 
of the distributions\cite{csvz}, 
and have been recently reconsidered by 
applying other numerical methods \cite{lubeck2,cvz,romu2}. 
From the theoretical standpoint it is
very surprising that small modifications
of the microscopic dynamics would lead to different universality class. 
However, no analytical demonstration of distinct universality
classes in sandpiles has been presented up to now. On the contrary, 
many theoretical arguments in favor of  a single universality class 
can be found in the literature\cite{rg}.

In Table~\ref{exps} we summarize the 
critical exponents found for each model. The quoted values indicate,
beyond numerical uncertainties, that the models 
discussed here belong to three distinct universality classes. 
Striking differences appear between the BTW and the Manna model. 
Beyond the numerical values of critical exponents, we observe for the first 
time the lack of self-averaging in the BTW FES. This property 
is related to its deterministic dynamics, and finds consistent 
analogies in the waves of toppling description\cite{topple}. 
The lack of self-averaging 
could also be the origin of  the multiscaling features  
recently observed by De Menech et al.\cite{stella} in the driven BTW sandpile. 
From this discussion it appears that the introduction of stochasticity
is a relevant modification for the critical behavior. 
At this point it is worth noting that the Manna model has been considered 
for a long time as a non-Abelian model. The opposite has been pointed out 
recently by Dhar\cite{mannadhar}, by means of rigorous arguments. 
The conjecture 
that Manna and BTW sandpiles belong to different universality classes because 
the former is non-Abelian has then to be abandoned. 
Stochasticity {\it per se}, however,  does not define a unique universality 
class, as evidenced by the distinct critical properties of  the Manna 
and shuffling FES models. The origin of the different behavior can be 
traced to the nonlocal nature of the shuffling model dynamics, as 
we shall make clear later. 

In summary, our numerical results are in good agreement with
the most recent measurements of driven sandpiles, confirming that 
the two cases share the same critical behavior. In addition, the
FES framework enlarges the set of exponents that can be measured,
providing new tools for the characterization of critical behavior and 
universality classes  in different models. 
 
\subsection{Avalanche and spreading exponents}

In order to compare the exponents found in fixed-energy simulations with
the usual avalanche exponents $\tau_s$ and $\tau_t$,
we relied on scaling relations. However, avalanches can also be studied 
in the FES case, in simulations of critical ``spreading''.  
Let us first define what is a spreading experiment in a 
system with an absorbing-state\cite{reviews}. 
In such experiments, a small perturbation (a single active site, 
for instance) is created in an otherwise frozen (absorbing) configuration. 
In the supercritical regime, the ensuing activity has a 
finite probability to survive 
indefinitely, reaching the stationary state deep inside 
the (growing) active region. In the subcritical regime, 
activity will decay exponentially. In each spreading sequence, it is 
customary to measure the spatially integrated activity $N(t)$, 
averaged over all runs,  and the 
survival probability $P(t)$ after $t$ time steps. 
At the critical point separating the supercritical and subcritical regimes,
these quantities have a singular scaling: $N(t)\sim t^\eta$ and 
$P(t)\sim t^{-\delta}$, where $\eta$ and $\delta$ 
are called spreading exponents.
If we can define the substrate over which the  activity spreads uniquely,
this spread of activity is the same as an avalanche 
in a sandpile model\cite{mdvz}. 

Sandpile models, however, have 
infinitely many absorbing configurations. In the infinite-size limit, an infinite
number of energy landscapes correspond to the same value $\zeta$. 
(For real-valued energies, as in the shuffling 
model, this infinite degeneracy already appears for finite systems.)
In this case spreading properties at a given value of the control parameter
$\zeta$ will depend on the initial configuration in which the system is 
prepared. It is even possible to observe nonuniversality 
in the spreading exponents, a feature that sandpiles share with the
pair contact process (PCP)\cite{dnu,pcp} and other systems with 
infinitely many absorbing configurations\cite{mendes,many,JSP}. 

In order to have well defined spreading exponents 
(that can be related to the avalanche
exponents of a driven sandpile), we have to 
define uniquely the properties of the energy landscape for spreading 
experiments.
One possibility is to use the absorbing
configurations generated by the fixed-energy 
sandpile itself for initial configurations. 
Suppose we use such a 
configuration for a spreading experiment,
by introducing an active site. Repeating  this 
process many times, we obtain the spreading properties for so-called 
``natural absorbing configurations''\cite{reviews}. 
A second option is to use
the substrate left by each spreading process as the initial condition
for the subsequent one. After a transient time the system
will flow to a stationary state with well defined properties, in which 
each initial configuration is a ``natural configuration''. 
On the other hand, this second definition of a 
spreading experiment is identical
to slow driving, except that energy  is strictly conserved
(the active site must be  generated by a mechanism that does not change
the energy)\cite{cmv}. 

By performing spreading experiments close to $\zeta_c$, it is possible 
to obtain directly the avalanche and spreading scaling behavior, as well
as the divergence of characteristic 
lengths approaching the critical energy. A
preliminary study in this direction for the BTW model confirms the 
uniqueness of the critical behavior at $\zeta_c$\cite{cmv}. 
Interesting results have also been obtained for 
the spreading properties in a FES  
mean field model\cite{china}. A more complete study of 
spreading exponents in a variety of sandpile models is a promising path 
toward the complete characterization of their critical behavior.
 
\subsection{Comparison with theoretical results}

In earlier sections we presented two alternative theoretical 
descriptions for sandpile models. We compare our numerical results with 
theoretical predictions in order to assess the validity 
of these theoretical frameworks,
and the eventual improvements needed for a complete 
description of sandpile models.

In Sec. III we introduced a Langevin description 
that takes into account the absorbing nature of 
the phase transition in FES models. 
Unfortunately, a rigorous derivation of the noise terms has 
not yet been made. The assumption of RFT-like noise terms 
leads to the Langevin 
description of Eq.~(\ref{main}). This differs from 
the standard DP Langevin description for the presence of a 
non-Markovian term. Only in the case that this term is irrelevant 
the theory belongs to the universality class of RFT. 
From a physical point of 
view this means that the local 
coupling between the activity field $\rho_a({\bf x},t)$ 
and the energy field $\zeta({\bf x},t)$ is 
irrelevant on large scales. 
In other words the activity spreads on an effective average 
energy substrate whose only role is to tune the spreading probability.
This is indeed the same as a DP problem in which the critical 
parameter is tuned via the average energy $\zeta$. 

Casting a glance at our numerical results, the only model that has exponents 
compatible with the DP universality class is the shuffling FES. 
This is not unexpected; the model was indeed proposed by Maslov 
and Zhang\cite{zhang2} as a sandpile realization of directed percolation. 
At the basis 
of this behavior is nonlocal energy transport. 
As we emphasized 
in Sec. II, the shuffling model allows the transfer 
of the same parcel of energy several
times in the same time step. This introduces, on average, a strong mixing 
effect that makes energy diffusion slower. In this way the spread of activity 
is  effectively decoupled from the local 
fluctuations that the activity itself generates 
in  the energy field. 
On the other hand, Maslov and Zhang \cite{zhang2} noted
that in $d=1$, the nonlocal energy mixing is not capable of
destroying correlations and, following a transient, 
the model exhibits non-DP scaling. While the exponents summarized 
in Tab.~\ref{exps}  are compatible with the DP 
universality class, we note that the 
dynamic scaling properties of the shuffling model 
show systematic biases that could signal 
a nonasymptotic behavior for some observables. 
We cannot therefore exclude completely that the model 
is still in a transient regime, that could finally lead 
to a different critical behavior, as happens in $d=1$. 

The Manna and BTW FES models, by contrast, exhibit 
critical exponents different from those of DP. In these models, the energy 
redistribution during toppling is strictly
local, and the spread of activity is always correlated 
with the energy fluctuations generated during toppling processes. 
It is then  reasonable to expect that a Langevin theory has
to take into account fully the non-Markovian 
term. It may be also possible to derive the pertinent stochastic equations
and the noise correlations applying more rigorous treatments, as 
in Ref~\cite{rdex}.

The moving interface picture is also afflicted by our ignorance 
of the correlations between the quenched noise terms 
appearing in the equations 
(see Sec. IV). By suitable approximations it has been shown that the Manna 
model could belong to the LIM universality class. Our numerical results 
show that the stationary 
critical properties are compatible with this universality class. The dynamic
properties, however, show anomalies that are not compatible 
with LIM. The origin of these anomalies deserves 
a more accurate analysis, and might be understood if we had a better
grasp of the noise terms in the interface representation.
It is interesting, in this context, that the BTW model, 
for which the mapping to the
interface representation seems most straightforward, 
defines a universality class {\it per se},
incompatible with linear interface depinning with columnar disorder. This is 
probably due to the strong nonlinearity introduced by the local velocity 
constraint implicit in the $\Theta$-function of Eq.~(\ref{hdyn}).

While neither theoretical approach allows an exact 
characterization of sandpile models,
they appear to be conceptually very relevant, because they provide an 
answer to the basic questions of why driven sandpile models show SOC. 
The genesis of self-organized criticality in sandpiles is a 
continuous absorbing-state 
phase transition. The sandpile exhibiting the latter may 
be continuous or discrete, deterministic or stochastic.  To transform the 
conventional nonequilibrium phase transition to SOC, we couple 
the local dynamics of the sandpile to a ``drive" 
(a source with rate $h$).  
The relevant parameter(s) \{$\zeta$\} associated 
with the phase transition are controlled 
by the drive, {\it in a way that does not make explicit reference} 
to \{$\zeta$\}.  
Such a transformation involves slow driving ($h \to 0$), 
in which the interaction 
with the environment is contingent on the presence or 
absence of activity in the system 
(linked to \{$\zeta$\} via the absorbing-state phase transition). 
Viewed in this light, ``self-organized criticality" 
refers neither to spontaneous or 
parameter-free criticality, nor to self-tuning. 
It becomes, rather, a useful concept for describing systems that, in
isolation, would manifest a phase transition between active 
and frozen regimes, and that are in fact driven slowly from outside.

A second class of theoretical questions concern the critical 
behavior (exponents, scaling functions, power-spectra, etc.) 
of specific models, and whether these can be grouped 
into universality classes, 
as for conventional phase transitions both in and out of equilibrium.
In this respect, the theoretical approaches presented here 
show a very promising path of improvements and modifications 
that could lead to the solution of many of these questions.

\section{Summary}

We studied three fixed-energy sandpile models, whose local dynamics are those of the 
BTW, Manna, and shuffling sandpiles, studied heretofore under external driving.  
The former two models are Abelian, the latter two
stochastic.  The results of extensive simulations, which are in good agreement
(via scaling laws), with previous studies of driven sandpiles, place the three models in
distinct universality classes.  Results for the Manna FES are consistent with the
universality class of linear interface depinning, while the
shuffling FES appears to follow directed percolation scaling.  Both these assignments,
however, are somewhat provisional, due to dynamic anomalies and apparent
strong finite-size effects.  The case of the BTW FES, which appears to define a
new universality class, is further complicated by violations of simple scaling
and lack of ergodicity.  
Examining the field-theoretic and interface-height descriptions
of sandpiles in light of our simulation results, we find 
that a more rigorous description of
noise correlations will be required, for these approaches 
to become reliable predictive tools.
Our results strongly suggest that there are at least 
three distinct universality classes
for sandpiles.  Whether others can be identified, and 
how the various classes can
be accommodated in a unified field-theoretic 
description, are challenging issues for
future study.

\vspace{1em}


{\bf ACKNOWLEDGEMENTS}
We thank M. Alava and  R. Pastor-Satorras for the many 
results on SOC they have discussed and shared with us  prior
to publication. We are also indebted with P. Grassberger for comments
and private communications. 
We also acknowledge A. Barrat, A. Chessa, D. Dhar,
K.B. Lauritsen, E.Marinari, L. Pietronero and A. Stella 
for very useful discussions and comments. 
M.A.M., A.V. and S.Z. acknowledge partial support
from the European Network contract ERBFMRXCT980183; M.A.M acknowledges 
also partial support from the Spanish DGESIC project PB97-0842, and 
Junta de Andaluc{\'\i}a project FQM-165.  R.D. acknowledges CNPq and CAPES
for support of computing facilities.



\begin{table}
\caption{\sf Critical exponents for the FES models studied here compared with
known values for DP and the LIM model \protect\cite{lesch}.
Figures in parentheses denote statistical uncertainties. 
}
\begin{center}
\begin{tabular}{|r|l|l|l|l|l|l|} 
model  & $\beta$ & $\beta/\nu_{\perp} $ & $z=\nu_{||}/\nu_{\perp}$ &$\theta$ & $\alpha$ & 
$\beta_W$ \\
\hline
BTW       & $\simeq 0.7$ & 0.78(3)  & 1.665(20) & 0.41(1)   & 1.01(1) & 0.62(1)  \\
Manna  	  & 0.64(1)      & 0.78(2)  & 1.57(4)   & 0.42(1)   & 0.80(3) & 0.51(1)  \\
Shuffling & 0.60(2)      & 0.76(3)  & 1.71(5)& $\simeq 0.46$ & $\simeq 0.96$     & $\simeq 0.57$    \\
DP        & 0.583(4)    & 0.80(1)  & 1.766(2)  & 0.451(1)   & 0.97(1)  & 0.55(1)       \\
LIM       &  0.64(2)   & 0.80(4) &  1.56(6)   &  0.51(2)    &  0.75(2) & 0.48(1)   \\
\end{tabular}
\end{center}
\label{exps}
\end{table}

\newpage
{\bf Appendix: Mean-Field description}

We have devised mean-field approximations for fixed-energy
sandpiles at the one- and two-site levels.  
While the present mean-field theory has nothing useful to say
about critical behavior, it is nonetheless interesting that a simple analysis
can yield reasonable predictions for the order parameter and transition points.
Consider first the
one-site approximation for the BTW FES.
Let $\rho_z $ be the density of sites with energy $z $.
Each site receives a unit of energy at rate $n_a$, the number
of active neighbors.  In the one-site approximation
the sites are treated as statistically independent, so that 
in a homogeneous system, the rate of arrival of particles at any site is
$2d \rho_a$ where
$\rho_a \equiv \sum_{z \geq 2d} \rho_z $ is the density of
active sites.  In addition to receiving energy, sites with $z \geq z_{th} = 2d$ make a transition to
$z - z_{th}$ at unit rate.  Hence the mean-field equations are

\begin{equation}
\frac{d \rho_z}{d t} = 2d\rho_a (\rho_{z-1} - \rho_z)
		       + \rho_{z+2d} - \theta_{z\!-\!2d} \; \rho_z,
\label{sitemf}
\end{equation}
where $\theta_j = 0$ for $j <0$ and is unity otherwise, and
$\rho_{-1}$ (in the equation for $z=0$) is of course zero.

This set of equations satisfies probability conservation
($\sum_z \rho_z $ is constant), and 
conserves the mean energy $\zeta \equiv \sum_z  z \rho_z $.
We try to find a stationary solution 
by introducing the
simplifying assumption that for $z \geq 2d$, the distribution
follows an exponential decay:

\begin{equation}
\rho_z = \alpha^{z-2d} \rho_{2d}, \;\;\;\;\;\;\;\; z \geq 2d.
\label{exp}
\end{equation}
Under this assumption,
the active-site density $\rho_a = \rho_2/(1-\alpha)$ in one dimension.
$\rho_0$ can be eliminated using normalization:

\begin{equation}
\rho_0 = 1 - \rho_1 - \frac{\rho_2}{1-\alpha}.
\label{erho0}
\end{equation}
Then the mean-field equations become

\begin{equation}
\frac{d \rho_1}{d t} = \rho_2\left[\alpha +
\frac {2}{1-\alpha} \left( 1 -2\rho_1 - \frac{\rho_2}{1- \alpha}
\right) \right]
\label{mf1a}
\end{equation}

\begin{equation}
\frac{d \rho_2}{d t} = \rho_2\left[\alpha^2 - 1 +
\frac {2}{1-\alpha} (\rho_1 - \rho_2) \right]
\label{mf2a}
\end{equation}
and
\begin{equation}
\frac{d \rho_z}{d t} = \rho_2\left[\alpha^z - \alpha^{z-2} +
\frac {2\rho_2}{1-\alpha} (\alpha^{z-3} - \alpha^{z-2}) \right] \;,
\;\;\;\;\;\;\;\; z \geq 3.
\label{mfza}
\end{equation}
The last equation implies that in the stationary state
$\rho_2 = \frac {\alpha}{2} (1 -\alpha^2)$, and therefore
$\rho_a = \frac {\alpha}{2} (1 + \alpha) $.  From Eq. (\ref{mf2a})
we then have $\rho_1 = \frac {1}{2} (1-\alpha^2)$
in the stationary state.  Thus $\rho_2 = \alpha \rho_1$ and the
distribution is exponential starting with $\rho_1$.
One readily
verifies that the r.h.s. of
Eq. (\ref{mf1a}) is also zero for
the stationary values of $\rho_1$ and $\rho_2$ given above.

The mean energy is given by:

\begin{equation}
\zeta = \rho_1 + \rho_2 \sum_{n =0}^{\infty} (n+2) \alpha^n = \frac{1}{2}
\frac{1+\alpha}{1-\alpha},
\label{zeta}
\end{equation}
so that
\begin{equation}
\alpha = \frac {2 \zeta - 1}{2 \zeta + 1},
\label{alpha}
\end{equation}
which shows that the active stationary state exists only
for $\zeta > \zeta_c = 1/2$.  Below this value, $\rho_a = 0$
and $\rho_1 = \zeta$.
(We have also verified that this solution is stable for
$\zeta > 1/2$.)
 In the active stationary phase,
\begin{equation}
\rho_a = \frac {4\zeta}{(2\zeta +1)^2} (\zeta - \zeta_c),
\label{op}
\end{equation}
so the order parameter exponent $\beta = 1$, the
usual mean-field value for systems lacking ``up-down" symmetry.

The two-dimensional case is
only slightly more complicated;
the mft equations may be written so:

\begin{equation}
\frac{d \rho_z}{d t} = 4\rho_a (\rho_{z-1} - \rho_z) -\theta_{z-4} \rho_z
+ \rho_{z+4},
\label{btwz}
\end{equation}
where $\rho_a \equiv \sum_{z \geq 4} \rho_z $.  We now suppose that in
the active stationary state, $\rho_z = \alpha^{z-4} \rho_4 $ for
$z \geq 4$.  Then $\rho_a = \rho_4/(1-\alpha)$, and proceeding as in
the one-dimensional case, one finds the stationary solution:

\begin{equation}
\rho_z = \frac{1}{4} (1-\alpha^{z+1}) , \;\;\;\;\;\;\; z \leq 2,
\end{equation}
and
\begin{equation}
\rho_z = \frac{\alpha^{z-3}}{4} (1-\alpha^4) , \;\;\;\; z \geq 3.
\end{equation}
The mean energy is

\begin{equation}
\zeta = \frac{3-\alpha}{2 (1-\alpha )} ,  
\end{equation}
so that
\begin{equation}
\alpha = \frac{2\zeta - 3}{2\zeta - 1} ,  
\end{equation}
showing that $\zeta_c = 3/2$ for the BTW sandpile in 2-d, in
this approximation.

It is also possible to derive two-site mean-field equations
without much difficulty. Denote the
probability for a NN pair of sites to have energies $i$ and $j$
by $\rho_{i,j}$.  The gain term, or rate of transitions {\it into} the state $(i,j)$,
due to one of the sites toppling or gaining a unit of energy from a
neighbor, is

\begin{equation}
\frac {d \rho_{i,j}^+ }{dt} = \rho_{i-1,j+2d}
   + \rho_{j+2d,j-1}
   + (2d-1)\left[ \frac {\rho_{i,j-1}
     \rho_{j-1,a}}{\rho_{j-1}}
   + \frac {\rho_{a,i-1} \rho_{i-1,j}}{\rho_{i-1}}
     \right],
\label{gaindd}
\end{equation}
where $\rho_{i,a} \equiv \sum_{j \geq 2d} \rho(i,j) $, and
we have used the fact that in the pair approximation
(i) the $2d$ neighbors of a given site are mutually independent, and
(ii) if $l$ and $n$ are neighbors of site $m$, then the three-site
probability 
$P(z_l,z_m,z_n) = \rho_{z_l,z_m} \rho_{z_m,z_n}/\rho_{z_m}$.
(The one-site probabilities are given by
$\rho_{i} = \sum_{j} \rho_{i,j}$.)
Note that there is no gain term for $i=j=0$: we expect no such
pairs in the active stationary state.
Similarly, the loss term is
\begin{equation}
\frac {d \rho_{i,j}^-}{dt} = \rho_{i,j}
   \left[ \theta_{i-2d} + \theta_{j-2d}
   + (2d-1) \left( \frac {\rho_{j,a}}{\rho_j}
   + \frac {\rho_{a,i}}{\rho_i} \right) \right].
\label{lossdd}
\end{equation}
The two-site probabilities are governed by
\begin{equation}
\frac {d \rho_{i,j}}{dt} = (1 + \delta_{i,j})
   \left[ \frac {d \rho_{i,j}^+ }{dt} - 
	   \frac {d \rho_{i,j}^-}{dt} \right].
\label{pmft}
\end{equation}

In the absence of a simple ansatz for the solution
of these equations, we analyze them
numerically.  For this purpose we choose a cutoff
and set $\rho(i,j) \equiv 0 $ for $i$ or $j > n$, where
$n$ is sufficiently large (in the range 20 - 36, depending
on $\zeta$), that $\rho_j$ is completely negligible
for $j \approx n$.  The coupled equations are integrated using a
fourth-order Runge-Kutta routine, starting from a product
Poisson distribution, $\rho_{i,j} = \rho_i^0 \rho_j^0$, with
$\rho_j^0 = e^{-\zeta} \zeta^j /j!$  (We verified that the
location of transition points does not depend on the
form of the initial distribution.)

The pair mean-field equations for the
one-dimensional BTW model predict
a {\em first order} transition at $\zeta_c = 0.91652$.
The active site density
jumps from zero to about 0.052 at this point.
Simulations \cite{dvz} also show a first-order transition, but at
$\zeta_c = 1$, with the order parameter jumping to about 0.14.
The energy distribution predicted by pair mft is approximately
exponential for $z \geq 3$ or so.

\begin{figure}[bt]
\centerline{
        \epsfxsize=7.0cm
        \epsfbox{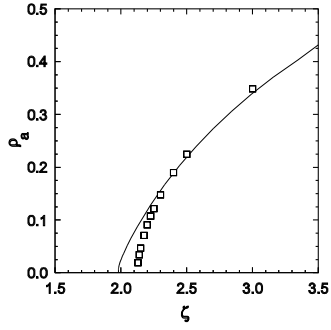}
        }
\caption{Pair mean-field prediction (line) and 
simulation results (squares) for
the active-site density in the two-dimensional BTW FES.}
\label{mffig1}
\end{figure}

In two dimensions, the pair approximation yields $\zeta_c = 1.98059$,
to be compared with the exact value of 2.125...
The transition is again discontinuous, but the jump in $\rho_a$
(from zero to about 0.0061), is very small.
(We find no evidence of a discontinuous transition in
simulations.)  At the critical
point, pair mft predicts $\rho_c = \rho_3 = 0.3328$, while simulation
yields $\rho_c = 0.434$. The energy
distribution decays exponentially for $z \geq 7$ or so.
Pair approximation and simulation results for the order parameter are
compared in Fig.~\ref{mffig1}.

The pair MFT is readily extended to the Manna FES defined in Sec. II.
In one dimension, when site i topples, the two particles are both sent to i-1
with probability 1/4 (similarly for site i+1), and with
probability 1/2, one each is sent to i-1 and i+1.
In the Manna sandpile some new
transitions, not allowed in the BTW model, make their appearance.
Enumerating the possibilities as above, one obtains,
for the one-dimensional exclusive Manna sandpile, the equations:
\begin{eqnarray}
\frac {d \rho_{\alpha \beta}}{dt} =
 & \;& \frac{1}{2} \left[\rho_{\alpha-1,\beta+2}
   +\rho_{\alpha+2,\beta-1}
   + \frac {\rho_{\alpha,\beta-1} \rho_{\beta-1,a}}{\rho_{\beta -1}}
   + \frac {\rho_{a,\alpha-1} \rho_{\alpha-1,\beta}}{\rho_{\alpha -1}}
   \right]  \nonumber \\
 & + & \frac{1}{4} \left[\rho_{\alpha+2,\beta} + \rho_{\alpha+2,\beta-2}
   + \rho_{\alpha,\beta+2} + \rho_{\alpha-2,\beta+2}
   + \frac {\rho_{\alpha,\beta-2} \rho_{\beta-2,a}}{\rho_{\beta -2}}
   + \frac {\rho_{a,\alpha-2} \rho_{\alpha-2,\beta}}{\rho_{\alpha -2}}
   \right]  \nonumber \\
 & - & \rho_{\alpha,\beta}
   \left[ \theta_{\alpha-2} + \theta_{\beta -2}
   + \frac{3}{4} \left( \frac {\rho_{\beta,a}}{\rho_{\beta}}
   + \frac {\rho_{a,\alpha}}{\rho_{\alpha}} \right) \right].
\label{empmft}
\end{eqnarray}
These equations predict a
{\em continuous} transition at $\zeta_c = 0.7500$,
in fair agreement with simulation ($\zeta_c \simeq 0.949$ \cite{man1d}).  
A straightforward generalization to two dimensions
yields a continuous transition at $\zeta_c = 0.625$, about 13\% smaller than the value
found in simulations ($\zeta_c = 0.7169$).

\newpage


\begin{references}


\bibitem{btw}
	P. Bak, C. Tang and K. Wiesenfeld,
	Phys. Rev. Lett. {\bf 59}, 381 (1987);
	Phys. Rev. A {\bf 38}, 364 (1988).

\bibitem{manna}
	S. S. Manna,
	J. Phys. A {\bf 24}, L363 (1991).

\bibitem{zhang1}
 	Y.-C. Zhang, Phys. Rev. Lett. {\bf 63}, 470 (1989);
	L.Pietronero, P. Tartaglia and Y.-C. Zhang, 
	Physica A {\bf 173}, 129 (1991).

\bibitem{directed}
	D. Dhar and R. Ramaswamy,
        Phys. Rev. Lett. {\bf 63}, 1659 (1989).

\bibitem{tadic}
           B. Tadic and D. Dhar,
           Phys. Rev. Lett. {\bf 79}, 1519 (1997).

\bibitem{dhar} 
	D. Dhar, 
        Phys. Rev. Lett.{\bf 64}, 1613 (1990);
	S. N. Majumdar and D. Dhar,
	Physica A {\bf 185}, 129 (1992); for a review see
	also D. Dhar, cond-mat/9909009.

\bibitem{priezz}V. B. Priezzhev, J. Stat. Phys. {\bf 74}, 955 (1994);
	E. V. Ivashkevich, J. Phys. A {\bf 27}, 3643 (1994);
	E. V. Ivashkevich,D.V.Ktitarev and V. B. Priezzhev,
	Physica A {\bf 209}, 347 (1994). 


\bibitem{rg}
	L. Pietronero, A. Vespignani and S. Zapperi,
	Phys. Rev. Lett. {\bf 72}, 1690 (1994).

\bibitem{rg2}
	J. Hasty and K. Wiesenfeld, J. Stat. Phys. {\bf 86}, 1179 (1997).

\bibitem{rg3}
	A. D{\'\i}az-Guilera, Europhys. Lett. {\bf 26}, 177 (1994).	

\bibitem{priezzdc}
	V. B. Priezzhev, cond-mat/9904054.

\bibitem{hwa}
	T. Hwa and M. Kardar,
	Phys. Rev. A {\bf 45}, 7002 (1992).

\bibitem{grin}   
	G. Grinstein, in 
	{\it Scale Invariance, Interfaces and Nonequilibrium Dynamics}, 
	{\it NATO Advanced Study Institute, Series B: Physics},
	vol. 344, A. McKane et al., Eds. 
	(Plenum, New York, 1995).

\bibitem{sor95}
	D. Sornette, A. Johansen, and I. Dornic, 
	J. Phys. I (France){\bf 5}, 325 (1995).
 
\bibitem{vz}
	A. Vespignani and S. Zapperi,
	Phys. Rev. Lett. {\bf 78}, 4793 (1997);
	Phys. Rev. E {\bf 57}, 6345  (1998).

\bibitem{dvz}
	R. Dickman, A. Vespignani and S. Zapperi,
	Phys. Rev. E {\bf 57}, 5095 (1998).

\bibitem{vdmz} A. Vespignani, R. Dickman, M. A. Mu\~noz, and Stefano Zapperi, 
	Phys. Rev. Lett. {\bf 81}, 5676 (1998).

\bibitem{midd} 
	O. Narayan and A. A. Middleton, 
	Phys. Rev. B {\bf 49} 244 (1994).

\bibitem{pacz} 
	M. Paczuski and S. Boettcher,  
	Phys. Rev. Lett. {\bf 77}, 111 (1996)

\bibitem{lau}
	K. B. Lauritsen and M. Alava, 
	cond-mat/9903346.

\bibitem{ala}
	M. Alava and K. B. Lauritsen,
	cond-mat/0002406.

\bibitem{fast}
	Deviations from criticality with respect to 
	the driving field can be obtained in the case of 
	fast driving. See ref.\cite{hwa} and 
	A. Barrat, A. Vespignani and S. Zapperi,
	Phys. Rev. Lett. {\bf 83}, 1962 (1999).

\bibitem{zls}
	S. Zapperi, K. B. Lauritsen, and H. E. Stanley,	
	Phys. Rev. Lett. {\bf 75}, 4071 (1995).

\bibitem{cmv} 
	A. Chessa, E. Marinari and A. Vespignani,
	Phys. Rev. Lett. {\bf 80}, 4217 (1998).

\bibitem{mtak}
	The question of open versus closed models for SOC is also discussed in
	A. Montakhab and J. M. Carlson, 
	Phys. Rev. E {\bf 58}, 5608 (1998).

\bibitem{tb88}
	An early study of sandpiles varying the total energy can be found 
	in C. Tang and P. Bak, 
	Phys. Rev. Lett. {\bf 60}, 2347 (1988).

\bibitem{reviews}
	R. Dickman,
	in {\em Nonequilibrium Statistical Mechanics in One Dimension}
	V. Privman, Ed. (Cambridge University Press, Cambridge 1996);
        G. Grinstein and M. A. Mu\~noz, in 
	{\it Fourth Granada Lectures in Computational Physics}, 
	Ed. P. Garrido and J. Marro, 
	Lecture Notes in Physics, {\bf 493}, 223 
	(Springer-Verlag, Berlin, 1997).
	J. Marro and R. Dickman,
	{\em Nonequilibrium Phase Transitions in Lattice Models}
	(Cambridge University Press, Cambridge, 1999).

\bibitem{lesch} 
	See H. Leschhorn, T. Nattermann, S. Stepanow and L-H. Tang,
	Ann. Phys. {\bf 6}, 1 (1997), and references therein.
   
\bibitem{fisher} 
	O. Narayan and D. S. Fisher, 
	Phys. Rev. B {\bf 48} 7030 (1993).

\bibitem{inf} 
	Like any other statistical model, a fixed-energy sandpile exhibits
	critical singularities only in the infinite-size limit.  In this limit
	the activity density is strictly zero for $\zeta < \zeta_c$, and
	positive for $\zeta > \zeta_c$, ensuring the stated inequality
	for $d \zeta/dt$ in the slowly-driven system.

\bibitem{mannadhar}
	D.Dhar, Physica A {\bf 270}, 69 (1999).

\bibitem{zhang2}
	S. Maslov and Y-C. Zhang, 
	Physica A {\bf 223}, 1 (1996).

\bibitem{rft}
	J.L. Cardy and R.L. Sugar, J. Phys. A  {\bf 13}, L423 (1980);
	P. Grassberger, Z. Phys. B {\bf 47}, 365 (1982);
	H.K. Janssen, Z. Phys. B {\bf 42}, 151 (1981).

\bibitem{bs94}
	The connection with the RFT theory has been also discussed
	for the Bak-Sneppen SOC model. 
	S.Maslov, M. Paczuski and P. Bak, 
	Europhys. Lett. {\bf 27}, 97 (1994); 
	P. Grassberger, Phys. Lett. A {\bf 200}, 277 (1995).

\bibitem{grasma}
	P. Grassberger and S. S. Manna, 
	J. Phys. (France) {\bf 51}, 1077 (1990).

\bibitem{manna2}
	S. S. Manna,  
	J. Stat. Phys. {\bf 59}, 509 (1990).

\bibitem{lubeck1}
	S. L\"ubeck and K.D. Usadel, 
        Phys. Rev. E {\bf 55}, 4095 (1997);
        ibid. {\bf 56}, 5138 (1997).

\bibitem{stella}
	M. De Menech, A. L. Stella and C. Tebaldi, 
        Phys. Rev. E {\bf 58}, R2677 (1998);
	C. Tebaldi M. De Menech and A. L. Stella,
	Phys. Rev. Lett. {\bf 83}, 3952 (1999).

\bibitem{inex} 
	This is the {\it inclusive} version of the Manna model.
	Is it also possible to define an {\it exclusive} version in which the
	two toppling particles are forbidden 
	to go to the same neighbor site: H. Kobayashi and M. Katori,
	J. Phys. Soc. Jpn. {\bf 66}, 2367 (1997).        

\bibitem{ben}
	A. Ben-Hur and O. Biham, Phys. Rev. E {\bf 53}, R1317 (1996).

\bibitem{csvz}
	A. Chessa, H. E. Stanley, A. Vespignani and S. Zapperi,
           Phys. Rev. E {\bf 59}, R12 (1999).

\bibitem{lubeck2}
	S. L\"ubeck,Phys. Rev. E {\bf 61}, 204 (2000).

\bibitem{cvz}
	A. Chessa, A. Vespignani and S. Zapperi,
	Comput. Phys. Comm. {\bf 121-122}, 299 (1999).

\bibitem{romu}
	R. Pastor-Satorras, private communication.

\bibitem{test}
	P. Grassberger, private communication.

\bibitem{immortals}
	In the BTW model, one actually encounters {\it immortal}
	configurations --- in which activity never ceases ---
	having $\zeta < \zeta_c$.  The probability of generating
	such an initial configurations decays rapidly with system size,
	and in practice we have not seen them for $L \geq 160$.

\bibitem{dp}
	W. Kinzel,
 	 Z. Phys. {\bf B58}, 229 (1985).

\bibitem{cp}
	T. M. Ligget,
	{\it Interacting Particle Systems},
	(Springer Verlag, New York, 1985).

\bibitem{notarft1}
	While our ansatz simplifies the analysis of stationary states,
	a theory of transient or spreading dynamics may require retaining
	$\rho_c$ as an independent field, if its initial value
	differs from the stationary one, 
	$\rho_{c,st} = (1-\rho_{a,st})f(\zeta)$.  The situation is analogous
	to that of a `non-natural' initial density in the PCP \cite{JSP}.)

\bibitem{Bray}
        A. J. Bray,
        Adv. in Phys., {\bf 43}, 357 (1994).

\bibitem{cardy}
	This is the case of branching, 
	annihilating random walks with even numbers
	of offspring, also known as the ``parity conserving" 
	or ``directed Ising" 
	universality class.  See:
	P. Grassberger, F. Krause, and T. von der Twer, 
	J. Phys. A {\bf 17}, L105 (1984);
	P. Grassberger,
	{\it ibid}. {\bf 22}, L1103 (1989);
	H. Takayasu and A. Yu. Tretyakov, 
	Phys. Rev. Lett. {\bf 68}, 3060 (1992);
	I. Jensen, 
	Phys. Rev. E {\bf 50}, 3623 (1994);
	N. Menyhard and G. \'Odor,
	J. Phys. A {\bf 29}, 7739 (1996);
	J. Cardy and U. C. T\"auber,
	Phys. Rev. Lett. {\bf 77}, 4780 (1996).
	H. Hinrichsen,
	Phys. Rev. E {\bf 55}, 219 (1997);
	W. Hwang, S. Kwon, H. Park, and H. Park,
	Phys. Rev. E {\bf 57}, 6438 (1998).



\bibitem{future}
        M. A. Mu\~noz, R. Dickman, A. Vespignani and S.Zapperi,
        preprint (1999).        

\bibitem{noest}  
	A.J. Noest,
	Phys. Rev. Lett. {\bf 57}, 90 (1986).       

\bibitem{dcp1}  
	A.G. Moreira and R. Dickman,
	Phys. Rev. E {\bf 54}, R3090 (1996).

\bibitem{noest88}  
	A.J. Noest,
	Phys. Rev. B {\bf 38}, 2715 (1988).       

\bibitem{dcp2}
	R. Dickman and A.G. Moreira,
	Phys. Rev. E {\bf 57}, 1263 (1998).

\bibitem{madcp}  
	R. Calfiero, A. Gabrielli, and M. A. Mu\~noz, 
	Phys. Rev. E {\bf 57}, 5060 (1998).       

\bibitem{notarft2}
	In the former case, e.g., for a site-diluted contact 
	process \cite{dcp1},
	activity tends to be restricted to favorable regions (lower than
	average dilution).  In the present case, it is principally at the 
	boundaries between active and inactive regions that, as implied by
	the Laplacian in Eq. (\ref{conservation}), energy is transferred,
	and the effect is to move energy into the inactive region, thereby
	{\it enhancing the further spread of activity}.  
	Indeed, the simulations
	reported below reveal {\it none} of the hallmarks of quenched disorder
	in the contact process,
	such as logarithmic time-dependence in critical spreading, or
	generic power-law relaxation of temporal correlations
	\cite{dcp1,noest88,dcp2,madcp}.

\bibitem{janssen}
	H. K. Janssen,
	Phys. Rev. E {\bf 55}, 6253 (1997).

\bibitem{alonnote}
           Remarkably, the opposite identification is
           also possible in certain cases.  See
           U. Alon, M. R. Evans, H. Hinrichsen, and D. Mukamel,
           Phys. Rev. Lett. {\bf 76}, 2746 (1996).
                                                                          
\bibitem{nota_dis}
	Note that in the case of quenched {\em point} disorder
	the ``automaton'' dynamics (i.e., when the local velocity
	can only take the values $v=0,1$) is found 
	to be in the same universality class as the continuous
	equation \protect\cite{lesch}. 

\bibitem{barabasi}
        A. -L. Barab\'asi and H. E. Stanley,
        {\it Fractal Concepts in Surface Growth},
        (Cambridge University Press, Cambridge, 1995).

\bibitem{parisi}
	G. Parisi and L. Pietronero,
	Europhys. Lett. {\bf 16},  321 (1991);
	Physica A {\bf 179}, 16 (1991). 

\bibitem{nota_lau}
	It has been pointed out that the $\Theta$-function 
	leads to an additional effective noise term \protect\cite{lau},
	which could imply a different universality class
	for the automaton model and the continuous
	equation.

\bibitem{notalim1}
	Note that the quenched  disorder present in the LIM equations
	is mimicked in the RFT representation by the site-dependent
	non-Markovian term. The general equivalence between quenched 
	noise and non-Markovian evolution has been pointed out 
	in the context 
	of the so-called run-time-statistics; see M.Marsili,
	J. Stat. Phys. {\bf 77}, 733 (1994).

\bibitem{FV} 
	F. Family and T. Vicsek, J. Phys. A  {\bf 18}, L75 (1985).

\bibitem{juanma} 
	J. M. L\'opez, Cond-mat/9910167. To appear in Phys. Rev. Lett.
	See also J. Kertesz and D. E. Wolf, 
	Phys. Rev. Lett. {\bf 22}, 2571 (1989);
	 S. Das Sarma et al. Phys. Rev. E {\bf 53}, 359 (1996); 
	J. M. L\'opez and M. A. Rodr{\'\i}guez, 
	Phys. Rev. E {\bf 56}, 3993 (1997); J. Phys. I {\bf 7}, 1191 (1997).

\bibitem{arlim} 
	N-N. Pang and W-J. Tzeng, Phys. Rev. E {\bf 59}, 234 (1999).

\bibitem{romu2}
	R. Pastor-Satorras and A. Vespignani,
	cond-mat/9907307.

\bibitem{fss} 
	M. E. Fisher, in {\it Proceedings 
	of the International Summer School `Enrico
	Fermi', Course LI}, (Academic Press, New York, 1971);
	M. E. Fisher and M. N. Barber, 
	Phys. Rev. Lett. {\bf 28}, 1516 (1972).

\bibitem{man1d}
	R. Dickman, M. Alava, M. A. Mu\~noz, J. Peltola, A. Vespignani, and
	S. Zapperi, in preparation.

\bibitem{exact} 
	It is likely that, in fact, $17/8$ is the exact result, as can
	be verified by using Priezzhev's results \cite{priezz}.
	Grassberger, private communication.

\bibitem{annote}
	The strict linear relationship between $\rho_a$ and $\rho_c$
	supports our elimination of $\rho_c$ as an independent field,
	in the continuum description developed in Sec. III.

\bibitem{short} H. K. Janssen, B. Schaub and B. Schmittmann, Z. Phys. B
        {\bf 73}, 539 (1989).

\bibitem{unote}
	That $C(t)$ is insensitive to a change in the order of updating lends some
	support to the assertion that average properties are not strongly dependent
	on the kind of updating used for the BTW model.

\bibitem{topple}
        D. V. Ktitarev, S. Lubeck, P. Grassberger, and V. B. Priezzhev,
	cond-mat/9907157.

\bibitem{mdvz}
	M. A. Mu\~noz,  R. Dickman, A. Vespignani, and Stefano Zapperi, 
	Phys. Rev. E {\bf 59}, 6175 (1999). 
\bibitem{dnu}
	R. Dickman,
	cond-mat/9909347.

\bibitem{pcp}
	I. Jensen, Phys. Rev. Lett. {\bf 70}, 1465 (1993);
	I. Jensen and R. Dickman, Phys. Rev. E {\bf 48}, 1710 (1993).

\bibitem{mendes}
	J.F.F Mendes, R. Dickman, M. Henkel, and M.C. Marques,
	J. Phys. A {\bf 27}, 3019 (1994).

\bibitem{many}  M. A. Mu\~noz, G. Grinstein,
	R. Dickman and R. Livi,
	Phys. Rev. Lett. {\bf 76}, 451, (1996).   

\bibitem{JSP}  
	M. A. Mu\~noz, G. Grinstein, and
	R. Dickman, J. Stat. Phys. {\bf 91}, 541-569 (1998).    

\bibitem{china}
	S.-D. Zhang,
	Phys. Rev. E {\bf 60}, 259 (1999).

\bibitem{rdex} M. Doi, J. Phys. A{\bf 9}, 1465 (1976).
               L. Peliti, J. Physique {\bf 46}, 1469 (1985).
       	B. P. Lee and J. Cardy,
	J. Stat. Phys. A {\bf 80}, 971 (1995).

\end{references}
\end{document}